\documentclass[manuscript, sigconf]{acmart}

\AtBeginDocument{%
  \providecommand\BibTeX{{%
    \normalfont B\kern-0.5em{\scshape i\kern-0.25em b}\kern-0.8em\TeX}}}

\setcopyright{acmcopyright}
\copyrightyear{2022} 
\acmYear{2022} 
\setcopyright{rightsretained} 
\acmConference[IUI '22]{27th International Conference on Intelligent User Interfaces}{March 22--25, 2022}{Helsinki, Finland}
\acmBooktitle{27th International Conference on Intelligent User Interfaces (IUI '22), March 22--25, 2022, Helsinki, Finland}\acmDOI{10.1145/3490099.3511119}
\acmISBN{978-1-4503-9144-3/22/03}

\usepackage{tikz}
\usetikzlibrary{plotmarks}
\usepackage{graphicx}
\usepackage{color}
\usepackage{xspace}
\usepackage{xcolor}
\usepackage{multirow}
\usepackage{longtable}

\usepackage{amsmath}
\usepackage{caption}
\usepackage{subcaption}

\usepackage{enumitem}

\usepackage{tabularx}
\usepackage{array}

\newcommand{\para}[1]{\paragraph{#1}}

\begin{document}

\title{Investigating Explainability of Generative AI for Code through Scenario-based Design}

\author{Jiao Sun}
\authornote{Work done during the first author's internship at IBM Research AI.}
\affiliation{%
  \institution{University of Southern California}
  \city{Los Angeles}
  \country{USA}
}
\email{jiaosun@usc.edu}

\author{Q. Vera Liao}
\authornote{Work done while the second author was at IBM Research AI.}
\affiliation{%
  \institution{Microsoft Research}
  \city{Montréal}
  \country{Canada}}
\email{veraliao@microsoft.com}

\author{Michael Muller}
\affiliation{%
  \institution{IBM Research AI}
  \city{Yorktown Heights}
  \country{USA}}
\email{michael_muller@us.ibm.com}

\author{Mayank Agarwal}
\affiliation{%
  \institution{IBM Research AI}
  \city{Yorktown Heights}
  \country{USA}}
\email{Mayank.Agarwal@ibm.com}

\author{Stephanie Houde}
\affiliation{%
  \institution{IBM Research AI}
  \city{Yorktown Heights}
  \country{USA}}
\email{Stephanie.Houde@ibm.com}

\author{Kartik Talamadupula}
\affiliation{%
  \institution{IBM Research AI}
  \city{Yorktown Heights}
  \country{USA}}
\email{krtalamad@us.ibm.com}

\author{Justin D. Weisz}
\affiliation{%
  \institution{IBM Research AI}
  \city{Yorktown Heights}
  \country{USA}}
\email{jweisz@us.ibm.com}

\renewcommand{\shortauthors}{Jiao Sun et al.}

\begin{abstract}
What does it mean for a generative AI model to be \emph{explainable}? The emergent discipline of explainable AI (XAI) has made great strides in helping people understand discriminative models. Less attention has been paid to generative models that produce \emph{artifacts}, rather than \emph{decisions}, as output. Meanwhile, generative AI (GenAI) technologies are maturing and being applied to application domains such as software engineering. Using scenario-based design and question-driven XAI design approaches, we explore users' explainability needs for GenAI in three software engineering use cases: natural language to code, code translation, and code auto-completion. We conducted 9 workshops with 43 software engineers in which real examples from state-of-the-art generative AI models were used to elicit users' explainability needs. Drawing from prior work, we also propose 4 types of XAI features for GenAI for code and gathered additional design ideas from participants. Our work explores explainability needs for GenAI for code and demonstrates how human-centered approaches can drive the technical development of XAI in novel domains. 

\end{abstract}




\maketitle

\section{Introduction}


Generative AI (GenAI) is a class of machine learning (ML) algorithms that can learn from content such as text, images, and audio in order to generate new content. In contrast to discriminative ML algorithms, which learn decision boundaries, GenAI models produce artifacts as output, which can have a wide range of variety and complexity. One major recent development of GenAI is the introduction of OpenAI's GPT-3~\cite{Brown2020LanguageMA} model, which can generate human-like language output and has striking versatility. Other generative language models have emerged that focus on specific domains such as software engineering, implementing use cases of auto-completing code~\cite{chen2021evaluating, kim2021code}, translating code from one programming language to another~\cite{roziere2020unsupervised}, and converting natural language to code~\cite{feng2020codebert}. The industry has begun to use these models to support software engineering practices, with the most prominent example being GitHub CoPilot~\cite{web:copilot}, a GenAI-based co-programming tool.


As a novel technology applied to novel domains, there are many open questions to be answered for how to make GenAI more capable and user-friendly. One open question is how to enable \textit{explainability}---allowing users to understand and have a better mental model---of GenAI. Recent works by \citet{goodfellow2020generative} and \citet{ross2021evaluating} have explored developing more interpretable GenAI models that follow more human-understandable processes. However, a more comprehensive view of explainability for GenAI is lacking: what do users need to understand about a GenAI model in order to effectively achieve their goals when working with it? In this paper, we build a foundational understanding of explainability needs for GenAI in the context of generative code models.

This question of \emph{what do users need to understand about AI systems} is core to the nascent field of Human-Centered Explainable AI (HCXAI)~\cite{ehsan2020human, ehsan2021operationalizing,liao2021human}, which is a subset of the fields of human centered AI and human centered data science~\cite{aragon_human-centered_2022, aragon2016developing, kogan2020mapping, muller2019human, muller2020interrogating, muller2021hcai, geyer2021hai}.
Our work is informed by a few key lessons from recent work in HCXAI, mostly conducted in the context of discriminative ML (e.g., for decision-support systems). First, explainability needs should be considered broadly as \emph{any means of helping users achieve a better understanding of the AI system}. \citet{liao2020questioning} proposed to define user's explainablity needs by what questions they ask to understand the AI~\cite{liao2020questioning} and developed a framework of common questions. This framework demonstrates that users are interested in a broad range of explanatory information about an AI model, including its overall logic, how it reasons to produce a particular output, the training data, and its performance and range of output. However, user needs regarding \emph{generative} models were not explored in that work.

Second, XAI solutions that address explainability needs \emph{should not be limited to algorithmic explanations or showing model internals}. Depending on user needs, it may be more critical to provide transparent information about a model's capabilities, limitations (e.g. uncertainty~\cite{bhatt2021uncertainty}) or provenance~\cite{arnold2019factsheets}. Moreover, users may need additional information beyond algorithmic explanations to fill in gaps of understanding. For example, \citet{ehsan2021expanding} proposed that \textit{social transparency} -- making visible the socio-organizational factors that govern the use of AI -- can help users form a socially-situated understanding of an AI system and take more effective actions with it. 


Finally, perhaps the most important lesson from HCXAI is that users' explainability needs \emph{emerge in a usage context}, guided by their goals and shaped by their backgrounds, expectations, as well as socio, organizational and cultural contexts~\cite{liao2020questioning,liao2021human,ehsan2021pitfalls}. It is thus necessary to follow a user-centered approach to understand explainability needs by involving target users and leveraging HCI methods that allow inquiry within the context of usage.

Based on these lessons, we adopted a scenario-based design method 
by constructing realistic usage scenarios for three use cases of GenAI for code: code translation, code auto-completion, and natural language to code. We invited 43 software engineers to participate in 9 workshops to elicit their explainability needs and design ideas around these scenarios. We adapted the question-driven method of Liao et al.~\cite{liao2020questioning, liao2021question} to elicit and comprehensively explore participants' explainability needs by what kind questions they would ask in the scenarios. We also gathered feedback and design ideas from participants for four kinds of XAI features that we propose for the uses cases of GenAI for code. Our work makes three main contributions to the IUI community:

\vspace{-1mm}
\begin{enumerate}
    \item We identify 11 categories of explainability needs in the context of Generative AI (GenAI) for code, for which we provide definitions and examples. We further contrast these categories with previous XAI techniques for discriminative ML and discuss explainability needs unique to GenAI and code generation use cases. We believe we are among the first to explore users' explainability needs in an application domain of GenAI.

    \item We propose four kinds of XAI features to support users of GenAI for code, based on prior work and adapted to the domain of code generation. These features are: AI documentation, indications of model uncertainty, visualizations of model attention, and social transparency. Based on participants' responses, we provide concrete design recommendations to operationalize these features.
    
    \item Our work makes methodological contributions by combining scenario-based design, participatory design workshops, and a question-driven approach to elicit explainability needs. We also reflect on the values and limitations of this method to inform future work that explores GenAI in new domains.
    
\end{enumerate}
\section{Related Work}
We review three areas of related work that shaped our study: GenAI for code, explainable AI, and human-centered approaches to AI.

\subsection{Generative AI for Code}


The application of modern NLP techniques to programming language can be traced back to the {\em naturalness hypothesis}~\cite{devanbu2015new,hindle2016naturalness,allamanis2018survey}, that software is a form of human communication. This hypothesis opened the door for applying NLP techniques previously used on human natural languages to source code, and recent work in this space is summarized by \citet{talamadupula2021applied} and \citet{allamanis2018survey}. One example is how existing work on automatic, machine translation between human natural languages~\cite{nguyen2014migrating,oda2015learning} was applied to code. Specifically, the TransCoder~\cite{roziere2020unsupervised} system applied neural machine translation techniques to translate source code across different languages. Other GenAI models have been developed that implement other use cases, such as generating documentation for code~\cite{feng2020codebert}, auto-completing code~\cite{chen2021evaluating, kim2021code}, generating unit tests~\cite{tufano2020unit}, and finding duplicate code~\cite{guo2020graphcodebert}. Models trained on massive code data sets are even able to handle multiple use cases at the same time, such as PLBART~\cite{plbart}, CodeBERT~\cite{feng2020codebert}, GraphCodeBERT~\cite{guo2020graphcodebert}. Most recently, OpenAI released Codex~\cite{chen2021evaluating}, which is a GPT-based model trained on code from GitHub and powers their CoPilot~\cite{web:copilot} product. This model is capable of auto-completing code for various programming languages (e.g., Python, TypeScript, Go, Ruby), as well as generating code from natural language. The release of Copilot is seen as a revolution of AI-assisted software programming and has attracted much
attention since its release~\cite{cade21ai}.

Despite the fact that GenAI for code models still have room for improvement in the quality of their outputs -- e.g., TransCoder only produces a correct translation 30\%-70\% of the time depending on the source and target language~\cite{roziere2020unsupervised} -- recent work by \citet{weisz2021perfection} suggests that software engineers may nonetheless be tolerant of using such models in their work. Given the emerging productization of GenAI for code models, we believe it is necessary to develop a comprehensive understanding of the kinds of questions software engineers will have when working with such models to guide technical XAI work and design solutions to answer them.

\subsection{Explainable AI}







Explainable AI (XAI) has spurred great academic, industry, and public interests in the past few years, thanks to the popularity of inscrutable ``opaque-box'' ML models. Many XAI techniques have been developed for discriminative ML models, both through producing directly interpretable models~\cite{caruana2015intelligible,lakkaraju2016interpretable} and generating post-hoc explanations for a trained opaque-box model~\cite{Zhou2016LearningDF, Lei2016RationalizingNP,ribeiro2016should,lundberg2017unified}. These explanations take a variety of forms. For example, \textit{global} explanations provide an overview of the model logic, while \textit{local} explanations elucidate the rationale behind a particular output. A full review of XAI techniques is beyond the scope of this paper and can be found in many recent survey papers~\cite{survey,adadi2018peeking,guidotti2018survey,lipton2018mythos}. Our work is most closely informed by, and intends to bridge, the emerging topic of explainability for generative models, and the inter-disciplinary field of Human-Centered Explainable AI (HCXAI)~\cite{ehsan2020human,ehsan2021operationalizing,liao2021human}.



Compared to discriminative models, much less attention has been paid to developing XAI techniques for generative AI models. Some explored ways to make generative models more directly interpretable, which is often framed as making the representation learning of a GenAI model in its latent dimensions semantically meaningful so people can directly examine the model internals \cite{goodfellow2020generative}. For example, disentanglement is a technique that seeks mappings between high-dimensional inputs and low-dimensional representations such that representation dimensions correspond to the ground-truth factors that generated the data~\cite{Ridgeway2016ASO}. Accordingly, some proposed disentanglement measures~\cite{Chen2018IsolatingSO, Ridgeway2018LearningDD} as a way to evaluate a GenAI model's interpretability. A recent study by \citet{ross2021evaluating} proposed a user evaluation task to evaluate the human interpretability of generative models, by the ability for people to interactively modify representations to reconstruct target instances.

Others explored visualization approaches to present the representations to help users (often model developers) make sense of what the GenAI model has learned. For example, ~\citet{ross2021evaluating} use sliders to let users dynamically modify representation dimensions and see how corresponding instances change. Recent HCI works also explored the approach of ``explainability through interaction" for generative models, by allowing users to interact with the input or guiding output generation process~\cite{louie2020cococo, louie2020novice,zhang2021method,ross2021evaluating}. Through observing immediate feedback from changes to the model output, people can make better sense of how a generative model works. For example, Zhang and Banovic developed a system that allows users to interactively explore the output space of an image generative model to assess the model quality~\cite{zhang2021method}. 

These varied approaches suggest that explainability is still a less than well-defined notion for GenAI. Here we adopt a HCXAI position that explainability, or the effectiveness of explanation, should be defined as enabling people's understanding of the AI to achieve their goals~\cite{liao2021human}. With this human-centered definition, many have argued that it is necessary to provide transparent information beyond the model internals, such as its performance, limitations, training data, and development procedure, to enable a holistic understanding of the AI and more actionable insights~\cite{vaughan2020human,paez2019pragmatic,bhatt2020explainable,liao2020questioning}. As discussed, Liao et al. proposed to identify users' explainability needs by eliciting the questions they ask to understand the AI~\cite{liao2020questioning,liao2021question}. This method allows approaching XAI from the users' perspective and thoroughly identify transparent information they need to achieve an understanding necessary for their goals, which, in the context of GenAI for code, could be about optimizing for the usage of the AI system and overall productivity. Liao et al.'s work was based on prior HCI work that defines ``intelligibility types" of information for context-aware intelligent systems by prototypical questions such as Input, Output, Why, What-if, etc. ~\cite{lim2009and,lim2010toolkit}. It also draws on social science literature showing that people’s explanatory goals can be expressed in different kinds of questions~\cite{hilton1990conversational}. We build on this line of work and adapt the question-drive method to elicit users' explainability needs for GenAI for code.

\subsection{Human-centered approaches to AI}

The term Human-centered AI has emerged in many academic works and public discussions~\cite{lee2020human,shneiderman2020bridging,riedl2019human,ehsan2021operationalizing, muller2021hcai, geyer2021hai}. While definitions vary, human-centered approaches to AI aim to develop AI systems that serve the needs, improve the conditions, and align with the values of human stakeholders. Research begins to develop practical methods that can help achieve these goals. One set of methods gained much attention lately under the umbrella term of ``participatory machine learning"~\cite{lee2019webuildai,halfaker2020ores,vinodkumar2020participatory}. Built on the tradition of participatory design, participatory machine learning emphasizes involving stakeholders, especially affected marginalized groups, into the development process early on to shape the overall goals and design choices of ML systems. Some suggested that abstracting user values from their participatory input is especially effective to guide ML modeling choices such as defining its optimization functions, features, constraints, and so on~\cite{zhu2018value,liao2019enabling,muller2017exploring}. This notion of driving technical development based on insights from empirical user studies is also at the core of broad IUI research~\cite{amershi2014power}.

One challenge to involving users to shape the design and technical development of AI systems is that these systems often do not exist yet for people to experience and provide realistic feedback. This challenge can be tackled by human-centered methods that allow ``envisioning future use possibilities''. One effective method is scenario-based design (SBD)~\cite{rosson2009scenario}. SBD suspends the needs to define system operations
by using narrative descriptions of how a user uses a system to accomplish a task, allowing people to respond to concrete interactions. We chose to use SBD to explore GenAI for code use cases as most software engineers do not have experience with such technologies. SBD also allows us to explore XAI design without the constraint of current technical feasibility, as adopted by several prior XAI works~\cite{ehsan2021expanding,wolf2019explainability}.


\section{Methodology: Scenario-based design workshops}


\label{sec:workshop_format}


We conducted 9 semi-structured workshops, with 3-6 participants in each, which lasted for 60-70 minutes. Due to the impact of the global COVID-19 pandemic, participants joined the workshop remotely via a video conferencing tool. We also used Mural\footnote{https://www.mural.co/}, which provides visual workspaces for virtual collaboration. Each workshop was based on one of three use cases of GenAI for code: code translation, code autocompletion, and natural language to code. In the following, we first introduce the three use cases then describe the workshop in detail, then participants and analysis.
\subsection{Use Cases and Scenarios} 


We focus on three specific use cases of GenAI for code, based on the preferred choices that could deliver high value for software engineering tasks from a pre-study survey with 81 people who responded to our study recruitment message.
\begin{itemize}
    \item \textit{Code translation}, in which a generative model translates source code from one language (e.g., Java) to another (e.g., Python). This task has been an important benchmark for technical work in GenAI for code ~\cite{Lu2021CodeXGLUEAM} and has gained extensive attention from both industry and academia~\cite{feng2020codebert, guo2020graphcodebert, plbart, roziere2020unsupervised, Wang2021CodeT5IU}. Such technologies can significantly reduce the cost and expertise barriers for code modernization work, in which a legacy codebase is ported to a modern programming language.
    
    \item \textit{Code autocompletion}, in which a generative model takes comments and source code as input (e.g. a function specification and/or signature), and produces code as output (e.g. the implementation of the function). This use case can fulfill pervasive needs of software engineers to improve their productivity and efficiency. Notably, autocompletion is one of the primary functions of GitHub Copilot.

    \item \textit{Natural language to code}, in which a generative model takes natural language (e.g. ``change the color of the button to blue'') and produces code as output (e.g. \texttt{button.setColor (Color.blue)}). This use case is another function offered by GitHub Copilot and represents a promising way to reduce entry barriers to programming.
\end{itemize}

\begin{figure*}[tp]
    \centering
    \includegraphics[width=\linewidth]{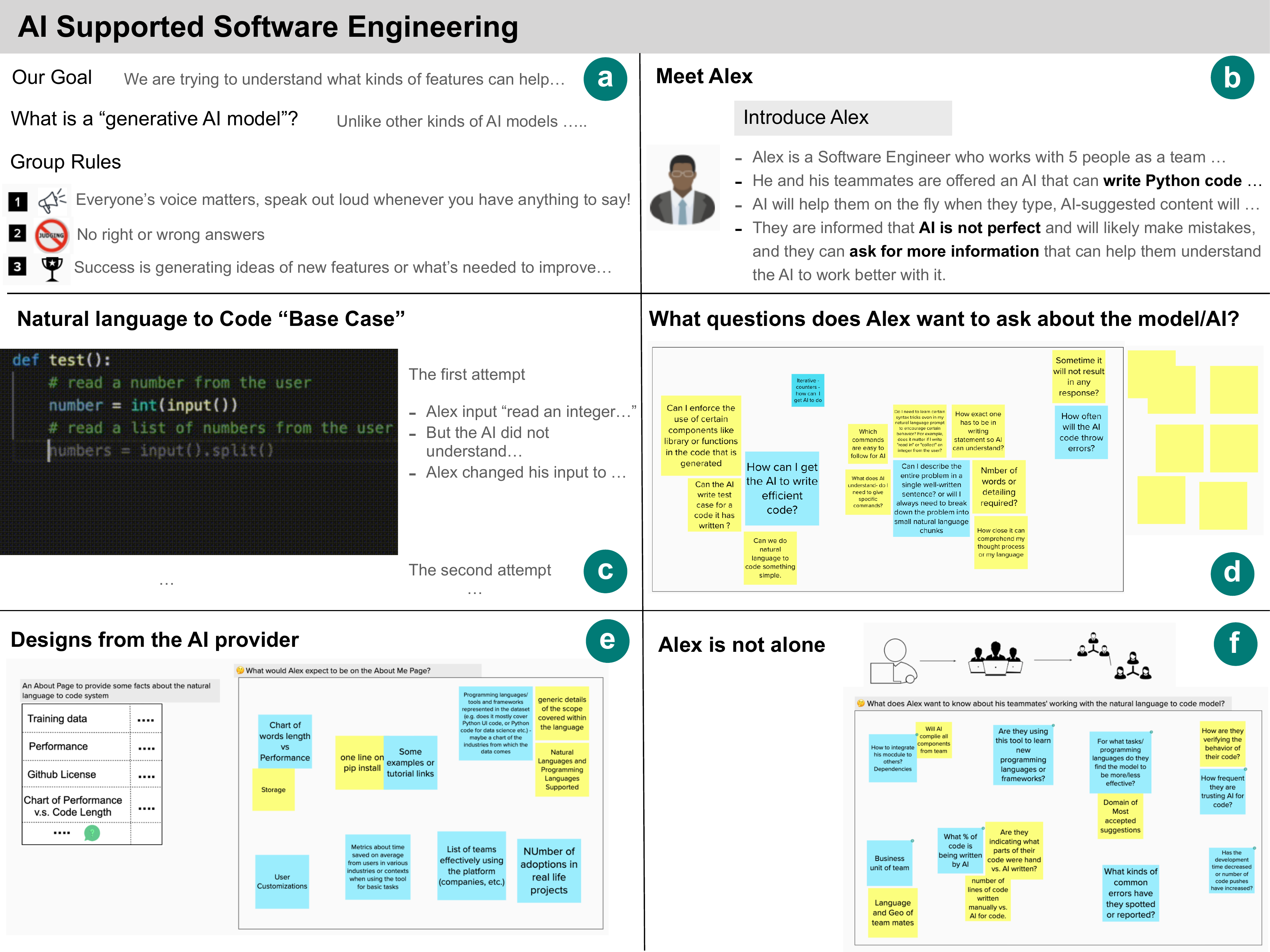}
    \caption{Overview of the Mural used in workshop \#6 (W6-NL2Code) for the code autocompletion use case. We introduce the workshop and set ground rules in \emph{(a)}, introduce the Alex persona in \emph{(b)}, and show the model output for the use case in \emph{(c)}. Then, we ask participants for what they want to know about the GenAI model in \emph{(d)}, elicit design ideas for AI documentation in \emph{(e)}, and ask participants to ideate on what information Alex would want to know about his team members' use of the model in \emph{(f)}. }
    \label{fig:workshop}
\end{figure*}

\begin{figure*}
\centering
\begin{subfigure}{.5\textwidth}
  \centering
  \includegraphics[width=0.5\linewidth]{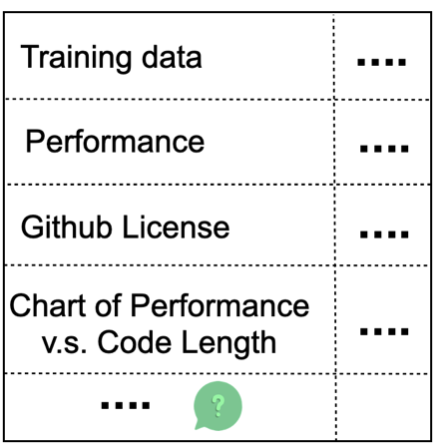}
  \caption{AI documentation}
  \label{fig:factsheet}
\end{subfigure}%
\begin{subfigure}{.5\textwidth}
  \centering
  \includegraphics[width=0.95\linewidth]{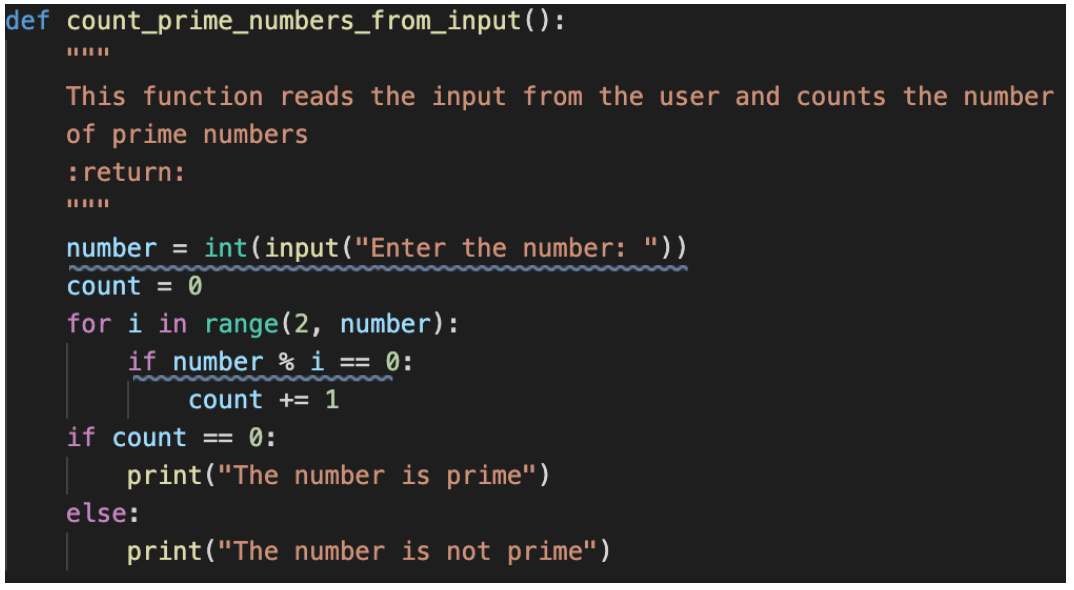}
  \caption{Uncertainty Indicator}
  \label{fig:bluelines}
\end{subfigure}%
\\
\begin{subfigure}{.5\textwidth}
  \centering
  \includegraphics[width=0.95\linewidth]{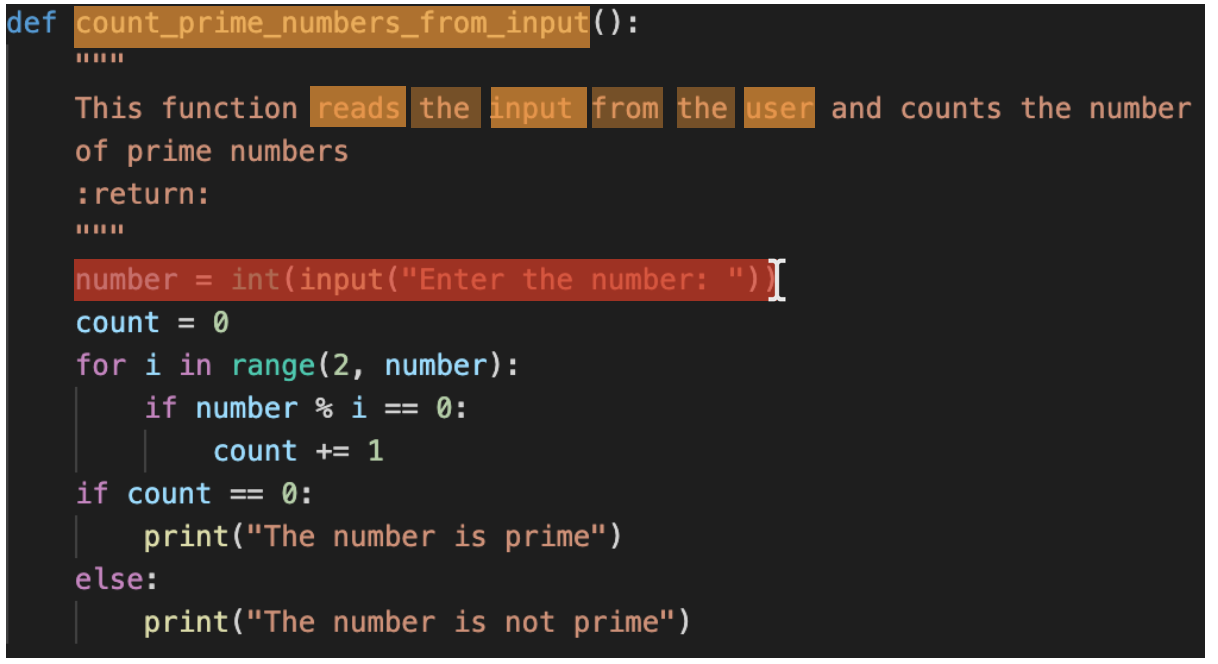}
  \caption{Attention Visualizer}
  \label{fig:highlight}
\end{subfigure}%
\begin{subfigure}{.5\textwidth}
  \centering
  \includegraphics[width=0.8\linewidth]{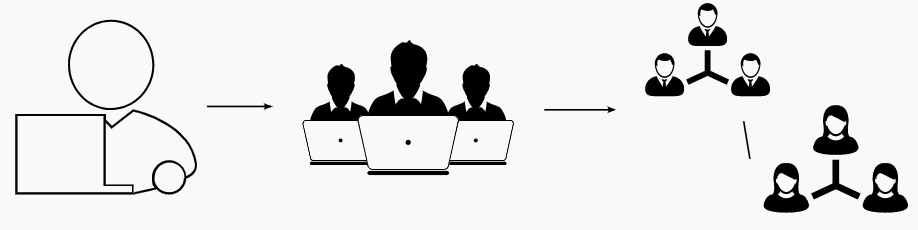}
  \caption{Social Transparency}
  \label{fig:social}
\end{subfigure}
\caption{Examples of UI probes used in the code translation use case. Probes used in the other use cases were similar in design but differed in the specific code examples used. Each design is translated from existing XAI approaches for discriminative AI. The probes were designed to elicit questions and spark discussion around Alex's information needs for working with a generative code model.}
\vspace{-2mm}
\label{fig:probes}
\end{figure*}


For each use case, we created a persona, Alex, who is a software engineer working in a team. In the scenario given to participants, Alex and his teammates are introduced to an AI system to support their work. They are told that they could ask for more information and new functions added to the AI system to help them understand and work better with the AI.  Figure~\ref{fig:workshop} $(b)$ shows the description of the persona for the code auto-completion use case.
A concrete scenario of a programming task that Alex would perform is shown in Figure~\ref{fig:workshop} $(c)$, and we show the programming tasks of the other two use cases (i.e., code translation and code autocompletion) in Appendix~\ref{app:other-two}.

 
To give participants a realistic experience, all AI-produced code in our scenarios was generated using state-of-the-art generative code models. We used TransCoder~\cite{roziere2020unsupervised} for the code translation use case and Copilot~\cite{web:copilot} for the other two use cases. For the code translation use case, we selected a programming solution to the problem of converting integers to their Roman numeral representations.\footnote{We selected a Java implementation of this problem from \url{https://algorithms.tutorialhorizon.com}.} One reason we selected this code example is that there was a subtle bug in the translation generated by TransCoder~\cite{roziere2020unsupervised}. This bug was pointed out by the workshop facilitator when introducing the scenario, to allow us to probe on participants' reactions to the limitations of the AI. For the code autocompletion and natural language to code use cases, we selected a programming solution to the problem of counting the number of prime numbers from a given input. We sampled this problem from Project CodeNet~\cite{codenet}, a large dataset of code samples. Based on CodeNet's metadata, the acceptance rate of Python solutions for this problem is lower than 50\%, indicating that it is a non-trivial coding problem for which GenAI models might provide assistance.
 


\subsection{Workshop Format and Procedure}
In each workshop, the facilitator first stated the purpose of the workshop, introduced concepts about generative AI for code, and asked for participants' verbal consent and permission to record. Then, the facilitator started the recording and introduced the Alex persona and one of the programming problem scenarios described above. The main part of the workshop was carried out in two stages: first, an open-ended question elicitation exercise to understand participants' explainability needs in the given scenario; and second, an ideation session in which 4 types of XAI features were used as design probes to elicit feedback and design ideas, to be described below. The content of the persona, scenario, and design probes were customized based on which use case was selected for the workshop. All discussions and screen activities were recorded, transcribed, and analyzed, together with the content on the Mural pages.

\subsubsection{Stage 1: Question elicitation exercise}
The first stage of the workshop was designed to elicit what kinds of questions participants would have for understanding the GenAI in the given scenario, based on the question-driven approach to XAI design proposed by Liao et al.~\cite{liao2020questioning,liao2021question}. After describing Alex's persona and the task scenario, the workshop facilitator asked participants: \emph{``Put yourself in Alex's shoes, what does Alex want to know? What questions does Alex want to ask about the natural language to code model/AI?''} Participants were given several minutes to come up as many questions they could, posted as sticky notes on the Mural page. Next, the facilitator worked with participants to collaboratively cluster similar questions together. Participants were asked to speak out loud as they moved around the sticky notes to describe their thought process and rationale. This process encouraged participants to read each other's questions, ask for clarifications, and have discussions. After the clustering, the facilitator chose one sticky note in each cluster to represent the cluster, then invited participants to cast up to 3 votes for questions they considered to be the most important to address. After the voting session, participants were asked to share their rationales behind the votes to exchange ideas and stimulate new ideas.

\subsubsection{Stage 2: Ideation on XAI features} 
In addition to exploring general explainability needs, our study also aims to explore a few areas to develop XAI features that can address these needs. Given the novelty in both the technical XAI approaches for GenAI and the application domain, we adapted existing ideas from XAI solutions for discriminative ML models as a starting point of our discussions. \citet{liao2021question} provides a suggested mapping between prototypical user questions (e.g., why, performance, data, output) and XAI methods or features that can answer the questions, primarily in the context of discriminative AI for decision-support systems. We built upon this work, as well as other work that explores XAI or transparency features~\cite{ehsan2020human, guo2019visualizing, bhatt2021uncertainty},  to select and adapt features that could apply to the context of GenAI for code. We also took into consideration their technical feasibility and potential values they can provide for the use cases. Through this process, we arrived at the following XAI features:
\begin{itemize}
    \item \textit{AI documentation} is embodied by the recently-proposed concepts of FactSheets~\cite{arnold2019factsheets, richards2020methodology}, Model Cards~\cite{wadhwani2020machine}, and About ML pages~\cite{raji2019ml}. AI documentation is developed by AI model or system providers to give users information about its purpose, performance, safety, and provenance. Liao et al. \cite{liao2020questioning} suggest that this type of feature could be used to address several categories of explainability needs, including data, output, performance, and how (global logic) questions. We designed a table with 4 categories of facts about the model, without providing specific information, as shown in Figure~\ref{fig:probes} (a). This table served as a probe to elicit other categories of information that should be included in AI documentation for generative code models. We intentionally left the content incomplete to inspire design ideation from participants.
    
    \item \textit{Uncertainty indicator} is a feature to communicate the model's uncertainty level about its output. Uncertainty is considered a critical form of transparency that can alert people about the limitations of AI~\cite{bhatt2021uncertainty}. In discriminative ML, uncertainty can take the form of a confidence score for a classification model or a prediction interval for a regression model~\cite{ghosh2021uncertainty}. In the context of generative code models, we envisioned that uncertainty could be assessed and represented at the line level. This observation was motivated by a recent work by \citet{Agarwal2020QualityE}, which demonstrates how to derive line-level confidences by aggregating token probabilities. In the design probe (Figure~\ref{fig:probes} (b)), we communicate the line-level lack-of-confidence with an wavy underline when it falls below some uncertainty threshold. 
    
    \item \textit{Attention visualization} is a type of local explanation to answer the \textit{why} question~\cite{liao2021introduction}: why did the model produce a particular output for a given input? For NLP tasks using deep neural networks, a common approach to explain a prediction is to utilize the attention mechanism, which provides a distribution over attended-to input units~\cite{wiegreffe2019attention} to indicate the relative importance of input units for the output. Prior work has shown how to use attention weights to visually depict which words in an input sequence, plus which words in previously-generated output, were responsible for a given token's output~\cite{Vig2019AMV, Vig2019AnalyzingTS}. In our design probe (Figure~\ref{fig:probes} (c)), we show the user querying a portion of output (in red) and seeing which tokens were responsible for that output (in yellow, where opacity is used to signal the strength of the attention weight). The opacities in the figure are based on the actual cross-attention weights from the model we used to generate the output.
    
    
    
    \item \textit{Social transparency} makes transparent the social contexts around the usage of an AI system to help people better understand the AI and how to utilize it. This feature was inspired by a recent work by \citet{ehsan2020human}, which examined the impact of making other users' past interactions with a decision-support AI system transparent. To explore social transparency in our context, we created an open-ended probe to invite ideation by visually emphasizing the fact that Alex is not working alone, but works within a group of other software engineers (Figure~\ref{fig:probes} (d)).
    
\end{itemize}

\vspace{-1mm}
\subsection{Participants}
\begin{table*}[]
\begin{tabular}{@{}lll@{}}
\toprule
\textbf{ID} & \textbf{Use Case} & \textbf{Participants}\\ \midrule
1 & Code Translation (CT) & Software Engineer (P16, P20), Researcher/Scientist (P5, P52) \\ \midrule
2 & Code Translation (CT) & \begin{tabular}[c]{@{}l@{}}Software Engineer (P3, P62), Hardware Verification Engineer (P54),\\Data Scientist (P54), Researcher/Scientist (P25)  \end{tabular}\\ \midrule
3 & Code Autocompletion (CA) & Software Engineer (P2, {\color{cyan} \ P53} , P63), Data Engineer (P64), Researcher/Scientist (P66) \\ \midrule
4 & Code Autocompletion (CA) & \begin{tabular}[c]{@{}l@{}} Software Engineer (P17, {\color{cyan} \ P39}), Researcher/Scientist (P37, {\color{cyan} \ P55}),\\ Bioinformatician (P48), Data Scientist (P61)\end{tabular}  \\ \midrule
5 & Code Translation (CT) & \begin{tabular}[c]{@{}l@{}}Researcher/Scientist (P19), Software Engineer (P46),\\Software Architect (P14, P45, P75), Data Scientist (P78) \end{tabular} \\ \midrule
6 & \begin{tabular}[c]{@{}l@{}}Natural language to code\\(NL2Code)\end{tabular}  & \begin{tabular}[c]{@{}l@{}}Software Engineer (P44), Researcher/Scientist (P71), Data Scientist (P80, P81)\end{tabular} \\ \midrule
7 & \begin{tabular}[c]{@{}l@{}}Natural language to code\\(NL2Code)\end{tabular} & \begin{tabular}[c]{@{}l@{}} Software Engineer (P59, P15), Software Architect (P41, P42),\\Researcher/Scientist (P13), Data Scientist (P30) \end{tabular} \\ \midrule
8 & \begin{tabular}[c]{@{}l@{}}Natural language to code\\(NL2Code)\end{tabular} & \begin{tabular}[c]{@{}l@{}}Software Engineer (P7, P22), Data Scientist (P8, P32)\end{tabular} \\ \midrule
9 & Code Autocompletion (CA) & Software Engineer (P57, P72), Data Engineer (P33)
\\ \bottomrule
\end{tabular}
\caption{Participants and their roles in our 9 brainstorming workshops about co-designing generative AI models for Code. Cyan indicates the participant does not have experience of working with AI. We will use abbreviations to refer to workshops throughout the paper, for example, W1-CT for workshop 1 about the CT (Code Translation) use case.}
\label{tab:workshops}
\end{table*}

\para{Recruitment.} We advertised our study within a large international information technology company, by distributing a survey targeted at software engineers working across many product lines and locations. 81 people responded. In the recruiting survey, we asked questions about their demographic information, self evaluation of programming skills and previous experience of working with AI. In addition, we asked them to rate their interest in the 6 candidate user cases of generative AI models for code (in addition to the 3 selected use cases we included test case generation, documentaion generation and code repair).

\para{Selection Criteria.} For \emph{code autocompletion} and \emph{natural language to code} use cases, we required participants to have 1+ year of Python programming experience. For the \emph{code translation} use case, we required participants to have 1+ year of both Python and Java programming experience. Among all sign-up responses, 56 (69\%) had at least 1+ year of Python programming experience. From this set, we prioritized to select those who have had experience with AI systems, and we further narrowed down to 43 participants based on their time availability.

\para{Demographics.} For the geographical location, the majority of our participants were from the United States (N=22; 65\%) and India (N=14; 41\%), with smaller numbers from Canada (N=3; 7\%), Germany (N=2; 5\%), Ireland (N=1; 2\%) and Brazil (N=1; 2\%). For the gender identity, 33 (77\%) participants identified 
as male, 8 (19\%) female, and 2 (5\%) preferred not to answer. Our participants also had diverse job roles: 17 (40\%) software engineers, 9 (21\%) researchers/scientists, 9 (21\%) data scientists, 5 (12\%) software architects, 1 data engineer, 1 bioinformatician and 1 hardware verification engineer. Among them, 40 (93\%) had experience working with AI in their jobs, while 3 of them did not. 

\para{Assignment.} We then assigned participants to 9 workshops (3 for each use case) as shown in Table~\ref{tab:workshops}, where we mark participants who do not have experience working with AI in blue.


\subsection{Analysis}

We conducted content analysis on the written content in Mural, and around 600 minutes of interviews were recorded and transcribed. In total, we extracted 487 messages from Mural and 249 paragraphs from the video transcript relevant to explainability. We followed the workshop structure, and coded the data in five parts: question elicitation exercise, discussions around AI documentation, uncertainty indicator, attention visualiser and social transparency.



To analyze the questions collected from the question elicitation exercise, we referred to prior work using prototypical questions to represent explainability needs by Liao et al.~\cite{liao2020questioning} in the context of discriminative ML systems, and ``intelligibility types''~\cite{lim2009and,lim2010toolkit} for context-aware intelligent systems by Lim and Dey,  and used these existing definitions to guide our coding. We also paid attention to new types of questions that were not covered in these prior works. Two researchers first independently coded 106 questions collected, and discussed their codes to reach a consensus. Then, they updated their coding independently with the agreed codes. As a result, they reached a high level of agreement on 66 out of 71 messages (93\%). After another iteration of discussion to finalize the coding schema, one of the two researchers coded the rest of questions. We discuss findings from this analysis in Section 4.


For the rest of the workshops, we coded participants' feedback and design ideas with regard to the four design probes. Our goal is to further understand users' explainability needs in the context of generative AI, as well as to inform concrete XAI designs that can address these needs. This part of coding was conducted by one researcher with frequent discussions with the other researchers. We discuss these findings in Section 5.

\section{Explanabilty needs for GenAI for Code}
\label{sec:general}



\begin{table*}[htp]
    \small
    \centering
    \begin{tabularx}{\linewidth}{XXp{1.6cm}}
        \toprule
         \textbf{Category \& Description} & \textbf{Exemplar Questions}\\
        \midrule
        \textbf{Input$^\ast$}. Questions about the kinds of input that the model can take or should be given.
        & \begin{tabular}[c]{@{}l@{}} - What kind of input can the AI reasonably generate code from?\\ - What types of [data types, data structures, algorithm...] does the AI\\ \hspace{1mm}  understand?\\ - How specific does the input need to be to get a good output?\\ - Is there certain thing I need to specify to get a good output?\end{tabular}
        \\

        \midrule
        \textbf{Output}. Questions about what the AI can produce, such as the output type, scope and system capabilities. 
        &
        \begin{tabular}[c]{@{}l@{}}- Will this produce idiomatic Python code, or a 1:1 translation?\\ - Will the translated code have the same [variable name, complexity, etc.] \\ \hspace{1mm} as the original? \\ - Will there be automated tests generated?\\ - Can the model generate code that does error/exception \\  \hspace{1mm}handling correctly?\end{tabular}
        &  
        \\
        \midrule
        \textbf{How (global)}. Questions about how the model operates to have a global understanding. 
        & 
         \begin{tabular}[c]{@{}l@{}} - How exactly does it generate code? \\ - How does the translation enforce the dynamic typing rules of Python?\\  - Can this AI use design patterns?  \\- Does the AI optimize for big O?\end{tabular}
         \\
        \midrule
        \textbf{Performance}. Questions about the quality of AI generated artifacts or the runtime performance (e.g. time taken) of the model.
        &\begin{tabular}[c]{@{}l@{}}- How correct is the translation guaranteed to be?\\  - How confident is the AI about the solution?\\ - How will the performance be for code of larger inputs? \\ - How much time will it take to translate the code?\end{tabular}
        \\
        \midrule
        \textbf{How to}. Questions about how to change the input to affect the quality or characteristics of the output
        & 
        \begin{tabular}[c]{@{}l@{}}- Can I give hints or specify my problem better to improve the model's \\ \hspace{1mm} output?\\- How can I get the AI to write more efficient code?\\ - How can I optimize for one type of output over another?\end{tabular}
        \\
        \midrule
        \textbf{Control$^\ast$}. Questions about options to customize or specify preferences for how the model should work. & 
        \begin{tabular}[c]{@{}l@{}}- Can I select a preference for [datatypes, packages, etc.] used?\\ - Can I tune the translation?\\ - Can I set specific regional settings?\end{tabular}
        \\
        \midrule
        \textbf{Why/Why not}. Questions about why the model produced a given output or why an input failed to produce the desired output. 
        &
        \begin{tabular}[c]{@{}l@{}}- Why didn't my input work? \\- How did the AI recognize the function from comments in this task?\\- Why did the AI think that its code satisfied the requirement? \\ - Why did that 4th attempt with the least programming thinking give a \\ \hspace{1mm} good result?\end{tabular}
        \\
        \midrule
        \textbf{Data}. Questions about the characteristics and provenance of the the data on which the model was trained.
        & \begin{tabular}[c]{@{}l@{}}- What data was this trained on? \\ - What kind of training set was the AI built with?\\ - Where does the code data the AI was trained on come from? \end{tabular}
        \\
        \midrule
        \textbf{System Requirements \& Impact$^\dagger$}. Questions about the requirements to use the system or its impact, to gauge the appropriate conditions of usage. & 
        \begin{tabular}[c]{@{}l@{}}- What are the hardware requirements for using the system?\\- Can I use the system in [a closed source project, my development \\\hspace{1mm} environment, etc]? \\
        - What is the energy consumption for using the model?\end{tabular}
        \\
        \midrule
        \textbf{Limitations$^\dagger$}. Questions about the limitations of the model's capabilities. & 
        \begin{tabular}[c]{@{}l@{}}- What limits are there to the model's function? \\ - What scenarios does it cover or not cover? \end{tabular}
        \\
        \midrule
                \textbf{What if}. Questions about what the output would be if the input changes or in hypothetical situations. & 
        \begin{tabular}[c]{@{}l@{}}- What to return if the input is invalid?\\ - What if I am translating from a language that does not allow overflow. \end{tabular}
        \\
        \bottomrule
    \end{tabularx}
    \caption{Question categories that emerged in our workshops. We applied the definitions of prototypical user questions in XAI Question Bank by Liao et al.~\cite{liao2020questioning} (in the context of discriminative ML, mainly for decision-support) and ``intelligibility types'' in Lim and Dey~\cite{lim2010toolkit} (for context-aware intelligent systems). Categories marked with a asterisk ($\ast$) are categories only appeared in Lim and Dey~\cite{lim2010toolkit} not Liao et al.~\cite{liao2020questioning}. Categories with
    dagger ($\dagger$) represent new categories emerged in our GAI for code context. For each question category, we provide the definition we used for coding and a few example questions asked by participants. } 
    \label{tab:question-bank}
\end{table*}


This section presents the results from the question elicitation exercise to understand what types of explainability needs particpants had for the GenAI for code use cases. As mentioned, we followed the definitions of prototypical user questions in XAI Question Bank by~\citet{liao2020questioning} and “intelligibility types” in~\citet{lim2010toolkit} to code the following categories: Input, Output, Performance, Data, How (global), Why/Why NOT, How to, What if, Control. We further identified two new categories that were not covered by the prior works: System Requirements and Impact, and Limitations. In Table~\ref{tab:question-bank}, we present these question categories, with their definitions, and example questions asked by participants. They are ordered by the frequency of being asked by participants. Categories unique to the GenAI for code are marked with a dagger ($\dagger$) sign. Categories that appeared in Lim and Dey~\cite{lim2010toolkit} but not Liao et al.~\cite{liao2020questioning} are marked with an asterisk ($\ast$) sign. Below we enumerate each question category and what we can learn about users' explainability needs with GenAI for code.

\para{Input.}\footnote{We adopt a similar definition of Input in Lim and Dey~\cite{lim2010toolkit}, as in what kind of inputs the model can take to generate code. Input was mentioned in Liao et al.~\cite{liao2020questioning} as training data went into the model. We consider that kind of question under the category of ``Data''.} To understand what kind of inputs the model can take or should be given was the most prominent explainability needs of participants, 
making up about 16\% of all questions. On one hand, participants wanted to have an \textit{overview} of the types or scope of inputs the AI can work with. Some asked about the AI's ability to process types of programming languages, data types, algorithms, language versions, and so on.  On the other hand, participants were eager to know how to \textit{optimize their inputs} to produce better outputs. For example, participants in W4-CA raised the question \emph{``Should Alex write smaller functions that are easier to name or extensive documentation for complex functions?''} and \emph{``Can I describe the entire problem in a single well-written sentence? or will I always need to break down the problem into small natural language chunks''} asked by W6-NL2Code.\footnote{Please see Table \ref{tab:workshops} for an explanation of the meaning of ''CA'' and ''NL2Code''. The prepended ''W'' number is the number of the workshop in the table. Also, we some of our transcribed comments came from the audio channel of workshop recordings, and we could not reliably determine who had spoken. Other comments recorded on sticky-notes could be attributed. Therefore, we report participants' comments in terms of which \textit{workshop} they occurred in, and we do not always know which \textit{participant} actually said them. We report participant IDs where we could determine them.} Participants' questions reflected their current mental models on how the GenAI works to process inputs, which were not necessarily accurate, suggesting that explanations may need to start from a high-level overview of how GenAI works in a specific code generation use case. It is interesting to note that this category is not a prominent need for decision-support AI using descriminative ML, since the input is often fixed or implicit. In contrast, the great variability and close coupling between input and output for GenAI made it a primary explainability needs to address.   





\para{Output.}  To understand what output the GenAI can produce is another frequent explainability need. Participants were mainly interested in the \textit{characteristics} of the output code, and the \textit{scope} of the output or system functions. Understanding the characteristics of the output can help users determine how to utilize the output, for example P5 from W1-CT commented: \emph{``you cannot ask Python to be as efficient as Java, but at the same time, Python may be more convenient than Java in some specific situations...''}. Understanding the characteristics can also guide users to assess the quality of output, identify potential shortcomings or errors for further actions, as commented by P19 from W4-CA: \emph{``if it's translating from a language like RUST that doesn't allow overflow, it's not possible into a language like Java allows it.''}. Meanwhile, some participants asked about the scope of the output, as to discover what the GenAI can do, such as whether it can generate test cases or multiple candidates as alternatives.

\para{How (global).} Participants also recognized the importance of having a a global understanding of the GenAI. Many wanted to have a high-level understanding of \textit{how the codes are generated}, such as \emph{``how exactly does it generate code? is it pulling from sites or using GANs to generate new code''} from W3-CA and \emph{``how the model comprehends the user input''} from W6-NL2Code. Some raised these questions based on their expectations of the input and output of the model (e.g., \emph{``Python is dynamically typed, Java is static, how do you enforce these typing rules?''} from W1-CT. Others inquired about how the model \textit{deals with specific types of input}, such as \emph{``How untagged text is used for analysis''} from W9-CA. These questions reflected interests in both a high-level description of how code generation works, and examining detailed model logic. For descriptive ML, global explanations often take the forms of showing how the model weighs different features or describing the general rules the predictions follow. How to communicate such a global view of model logic for GenAI is an open technical challenge. 



\para{Performance.} Many participants were concerned about understanding the performance, i.e. how well the GenAI works for the use case. These questions are critical for users to develop appropriate trust and adopt the system. We found the questions are mainly concerned about the \textit{overall performance}, \textit{the quality of a specific generated code} and the \textit{run-time efficiency} of the model  (e.g., inference time of the model, if the AI supports multi-threading). It is interesting to note that ``quality of code'' may be multi-faceted as participants mentioned correctness, runtime complexity, space complexity, understandability, and testability.  Many of these performance-related questions were asked to understand \textit{performance differences and limitations} with regard to different types of input (e.g. larger code). The interest in run-time performance is a unique need emerged in the GenAI for code context.

\para{How to.} These questions inquired about how to change or improve the input to get a better output. Answers to these questions can help participants choose strategies to make better use of the AI, such as 
\emph{``How to [make] requirement better [to improve the output]''?} from W9-CA and \emph{``How can I get the AI to write efficient code?''} from W6-NL2Code. 



\para{Control.} These questions were regarding options to customize or specify preferences for how the model should work. This category was not seen in Liao et al.'s work defining explainability needs for decision-support descriminative ML. Emergence of this type of question in our context suggests that users of GenAI for code are interested in having more control on the working of the model, as illustrated by P61 from W1-CT saying that \emph{``Oftentimes, you're not using the tool for just basic examples. I guess more in depth, like, explanations of what you can do with the model would be appreciated.''}



\para{Why/Why Not (local).} Different from asking \textit{How (global)} questions to understand the overall process or logic of the model, a \emph{Why (local)} question was asked to understand a specific output from the model. More often than not, the \textit{Why} question was triggered by a surprising or suspicious output, such as \emph{``Why did the AI think the code satisfied the requirement?''} at W7-NL2Code or \emph{``How did the AI recognize the function from comments in this task?''} at W4-CA. Sometimes the \textit{Why} question was asked in contrast to the expected output, and hence a \textit{Why Not} (the expected output) question, such as \emph{``Why didn't my input work?''} from W3-CA.



\para{Data.} Some participants brought up questions regarding the training data, especially regarding the provenance of the data, such as \emph{``Where does the code the AI was trained on come from?''} from W3-CA. Echoing findings in Liao et al.~\cite{liao2020questioning}, explanations on training data can help users gauge the capability of the model and its proper usage, by the validity of the training data and their alignment with ones' own programming tasks. How to define alignment of training data in the context of GenAI, however, remains an open question. 



\para{System Requirement and impact.} A new category emerged in our data that was not covered in prior works. This category is questions regarding requirements or impact of the system, which can help users gauge the appropriate conditions of usage. They were asked regarding the usage of the system instead of the underlying model. These questions were highly specific to the software engineer context, including compatible development environments, software and hardware requirements, dependencies, and so on. 


\para{Limitations.} Another category unique in our data is questions explicitly asking about the limitations of the model or system, such as what kind of scenarios it cannot cover or limitations to the model's functions. These questions emerged likely due to the complexity of the input and output spaces of GenAI, which is less straightforward to comprehend than a decision-support AI system. 


\para{What if.} Some asked about what the output would be given hypothetical changes to the input, which may further help participants understand how the AI makes decisions in a counterfactual manner, such as \emph{``what to return if the input is invalid?''} from W5-CT.

To summarize, we identified 11 categories of explainability needs in GenAI for code use cases. Four of these categories did not appear in Liao et al. as prominent needs for descriminative ML used for decision-support: Input, Control, System Requirements \& Impact and Limitations. The top five most frequent categories are Input, Output, How (global), Performance and How to. In Section 6, we will further discuss insights revealed about explainabilty needs that are unique to the GenAI technology and the novel use case of supporting code generation.


\section{XAI features for GenAI for Code}








        


As discussed in Section 3.2, we proposed four types of XAI features for GenAI for code and created design probes to elicit ideation from participants: AI documentation, uncertainty indicator, attention visualizer, and social transparency. In this section we discuss participants' response to derive concrete design recommendations.


\subsection{AI Documentation}

\begin{table*}[htp]
    \centering
    \begin{tabularx}{\linewidth}{>{\raggedright\arraybackslash}p{2.8cm}lX}
        \toprule
        \textbf{Category} & \textbf{Applied to} & \textbf{Definition} \\
        \midrule
        Examples \& Tutorials$^{\dagger}$ & Models & 
            Examples of input-output pairs generated by the model; Tutorials on how to use the model effectively, including what kinds of input can get high-quality outputs \\
        \midrule
        Software Engineering Capabilities$^*$ & Models & 
            Description of software engineering features or capabilities that the AI can support  (e.g., version information, stress testing for large loads, dependency handling, data structure, kernel status, coding style)
           \\
        \midrule
        Model Performance & Models & 
            Technical evaluation metrics of the generative model, including accuracy, performance, performance change by types of input, CPU consumption, and model inference time  
            \\
        \midrule
        Output Code\ Quality and Utility{$^{*\dagger}$} & Outputs & Metrics characterizing the generated code, including correctness, lint errors, code efficiency, time complexity; Metrics reflecting the system's impact on human productivity, including estimated time savings in conducting programming tasks, estimated improvements to code quality, comparisons with other kinds of GenAI for code tools \\
        \midrule
        Supported Languages \& Frameworks$^*$ & Models & 
            List of programming and/or human languages which the model is capable of understanding (e.g. as input) or producing (e.g. as output); List of programming frameworks or APIs which the model supports as input or output (e.g. React, Flask) \\
        \midrule
        Data  & Models  & 
            Information about what data the model was trained on, including its provenance and any applicable privacy policies, data usage guidelines, or code licences \\
        \midrule
        Control $^{\dagger}$ & Models & 
            Description of customization options or other mechanisms for users to control the output of the model; Description of how the model can be fine-tuned for additional use cases \\
        \midrule
        Deployment Requirements \& Platform$^*$ & Models & 
            Technical requirements for hosting the model, including: software dependencies, hardware requirements, cloud-hosting requirements, and supported IDE integration  \\
        \midrule
        Model Explanations & Models &
            Explanation of how the model operates (e.g. a basic description of how transformer models work or a visualization of model attention) \\
        \midrule
         Usage Rights &  Outputs & Information about usage restrictions and/or licensing terms for code produced by the model \\
         \midrule
        Optimal \& Poor Conditions & Models & 
            The conditions under which the AI model performs well or performs more poorly than expected \\
        \midrule
        Intended Usage & Models & 
            The use cases supported by the model; Other potential use cases that might be implemented via fine-tuning \\
        \bottomrule
    \end{tabularx}
    \caption{Categories of AI documentation for GenAI for code models and their outputs. Categories are sorted in descending order, based on the frequency of mention by participants. Categories marked with a $\dagger$ or $*$ have not been identified by previous works in AI documentation. We postulate that categories marked with a dagger ($\dagger$) emerged due to the context of \emph{generative} models, and categories marked with an asterisk ($*$) emerged due to the context of software engineering support. }
    \label{tab:document}
\end{table*}

AI documentation is advocated to increase the transparency and facilitate understanding of AI~\cite{Hind2019IncreasingTI, Hind2020ExperiencesWI, Richards2020AMF, Piorkowski2020TowardsEA, Piorkowski2021HowAD, Knowles2021TheSO}, and is embodied in recently proposed features of AI Factsheets~\cite{arnold2019factsheets, richards2020methodology} and Model Cards~\cite{Mitchell2019ModelCF}. 
However, what documentation should include for generative AI models is under-investigated, let alone GenAI for code use cases. We are interested in \textit{what categories of information about the model should be presented in AI documentation for GenAI for code}, and how they might differ from that of discriminative ML. We used a low-fidelity UI design of a factsheet with intentionally open-ended content, as shown in Figure~\ref{fig:factsheet},  and asked participants to brainstorm AI documentation categories that could help the user in the design scenario.

We summarize the categories identified from data analysis in Table~\ref{tab:document}, ranked by the frequencies of mentions by participants. In particular, we identified categories that surfaced in the context of GenAI for code that were not discussed in previous works on AI FactSheets or Model Cards: \emph{Examples and tutorials},
\emph{Software engineering capabilities}, \emph{Output code quality and utility},  \emph{Supported languages and frameworks}, \emph{Control} and \emph{Deployment Requirement \& Platform}. 

Consistent with findings from the question elicitation exercise in Section 5, participants expressed strong interests in understanding the scopes of inputs and outputs of the system. \emph{Examples and tutorials}, as the most frequently mentioned category, suggests that example-based explanations can be a fruitful area to explore for GenAI to help users understand the model inputs and outputs. Such interests are also expressed as requirements specific for software engineering tasks, including \emph{Software engineering capabilities} and \emph{Supported languages and frameworks}. Participants also wished to see information about \emph{performance}, \emph{data}, \emph{control}, \emph{system deployment requirement}, and \emph{global model explanations}, suggesting that providing documentation could help address corresponding explainability needs found in Section 5. 

Another interesting pattern is that participants requested categories of information regarding the generated code artefacts rather than the model itself (see the second column in  Table~\ref{tab:document}). Many demanded to see metrics about the \emph{code quality and utility}, which should be differentiated from model performance metrics such as accuracy. While the latter are commonly calculated with some ground-truth in a held-out test dataset, the former are concerned with characterizing the generated code and their utility for improving human productivity.


In general, documentation for GenAI for code should be transparent about the quality of the produced code from the software engineering perspective, and align with the professional culture and standards, such as communicating the license and regulatory requirements in both the training data acquisition and the usage of the produced code. Table~\ref{tab:document} is not meant to serve as an exhaustive template for AI documentation of generative AI models for code. Instead, we hope that these discovered categories can inspire what to document for generative AI models for code.

\subsection{Model Uncertainty}
Inspired by previous work on model uncertainty of generative AI models for code~\cite{weisz2021perfection} and communicating uncertainty of output for discriminative ML models~\cite{bhatt2021uncertainty,ghosh2021uncertainty}, we created the design probe in Figure~\ref{fig:bluelines} to communicate uncertainty for a line of AI generated code, indicated by a wavy underline. We asked participants to react to the feature and discuss \textit{how to improve the feature, including follow-up interactions, to better support users' understanding and their usage of the AI system}.

Their suggested improvement and follow-up interactions can be categorized as \textit{AI initiated} and \textit{human initiated} \cite{horvitz1999principles, parasuraman2000model}. For the AI to take the initiative, participants suggested to see more information from the model to help them understand the output and edit if necessary, for example to include suggesting alternative outputs, providing uncertainty explanations, and giving more fine-grained uncertainty. We elaborate on each category below.

\para{Alternative outputs}
Participants wanted transparency about what other alternatives the AI model had considered as output. Some participants had low expectations of generated code and hoped to gain insights from more options, as P63 from W3-CA stated that \emph{``I voted for the multiple [version] candidates, which I think is just generally helpful. Even if the top generated candidate was not quite right maybe one of the top five will be. Or will be better for whatever the particular use cases.''}. Some participants hoped to see alternatives with specific characteristics (e.g., better readability or optimization), as P80 from W6-NL2Code stated: \emph{``Obviously if human knows about the time complexity and the optimization of the code, he can fetch a better one. And at present, we do not know whether [the] AI [is] trying to be more on the readable side or more on the optimization side.''}

\para{Uncertainty explanation}
Participants desired to see \textit{uncertainty explanation} from the model, on why a particular part of the code was highlighted or why the AI was uncertain. This is an under-explored area in XAI. Participants suggested that the sources for uncertainty explanations could be what alternative options the model was struggling with (W8-NL2Code), or what goal or rationale the model follows (W6-NL2Code). We quote P71 from W6-NL2Code who proposed an interactive explanation to address this point: 
\begin{quote}
    \emph{``I've voted twice for an indication of what the model is trying to achieve, because I think that's very important. It could be very simple. It could be as simple as just reminding you exactly what natural language prompt this was derived from, so you can think to yourself: `what was it trying to do?' Uh, based on what I wrote, cause I know what I wrote, and I know what that means and then I can rewrite it from there or it could be more complex. It could, you know, if there's a serious natural language model behind this, maybe it's capable of generating speech back to me about what this code exactly does and then you can compare that yourself.''}
\end{quote} 

Another path to explain uncertainty suggested by participants was through \textit{highlighting corresponding parts} that contributed to the uncertainty. The corresponding parts could be from input, 
previously generated lines, or training data.

\para{Fine-grained uncertainty} 
Some participants wanted to see more than just a wavy line under outputs that are below an uncertainty threshold. For example, P71 from W6-NL2Code said: \emph{``...if there is some confidence percentage that could be helpful to, you know, like Green Zone, Red zone and Yellow zone etc., to give an idea of the highest priority I need to fix this.''} Related to the observation that code quality is multi-faceted, some participants wanted to understand the specific definition or aspect of uncertainty, such as whether it is regarding the correctness, time complexity, or runtime.




\vspace{1em}

Another area of suggested interactions requires the human to take the initiative, including allowing human input to resolve uncertainty and supporting interactive testing for the uncertain regions, as discussed below.

\para{Human input to resolve uncertainty}
A frequently requested interaction was that the AI waits for human input to resolve uncertainty. This could be realized by the AI prompting the human with a variety of questions, including confirmation (W3-CA), clarification (e.g., \emph{``show did you mean ... options''} from W1-CT), inspection (e.g., \emph{``Tell user to examine the line carefully and verify''} from W4-CA), preferences or how they would code  (e.g., \emph{``Ask user for decision making -- `do you want to keep this line or not' and highlight it with reasoning''} from W5-CT), and immediate modification (e.g., \emph{``Dialog box to type preferred translation (if Alex knows python'')} from W2-CT)).

\para{Interactive testing}
Another interaction that participants expected to perform is interactive testing for situations in which the model was uncertain, which can also facilitate understanding of the AI model. P81 from W6-NL2Code said that \emph{``... If we can probably give a similar interactive component [as Jupyter Notebook] where it gives you the option to fetch a context from the previous clients, and then run only that piece of code to see what the output is. Probably the person can make a better decision on whether to keep it or change.''}



The overall feedback from participants was that showing a line-level uncertainty indicator would be useful to guide users' attention to potentially low-quality code. However, showing uncertainty alone was not enough to satisfy their explainability needs. Many of the suggested interactions also serve the purpose of helping users better understand the model behaviors through interactive testing, alternative output it considered, and explaining the uncertainty. Through these interactions, users can develop a more appropriate mental model of the system that can help them interact more effectively in the long run.

\subsection{Attention Distribution}
Attention distribution is a popular approach to explain an individual prediction made by a neural networks model, by the relative importance of parts of input to determine the output \cite{Vig2019AMV, Vig2019AnalyzingTS}. Mapping to the context of  GenAI for code, we designed the UI in Figure~\ref{fig:highlight}, where the user can select a span of generated content. The AI can then highlight some spans in the previous content, to illustrate the idea of attention when making local decisions, indicating different attention weights via different opacity. 
We showed it to participants to gather their response on \textit{the utility of attention distribution visualizer for users of GenAI for code}.

In general, participants said that the attention features would be useful to help them understand how the model worked, and to guide them in modifying the input to produce better results. P42 from W7-NL2Code said that \emph{`` It's what trains me as the user to use the AI better''} and P63 from W3-CA said that \emph{``if you see an issue, you might be able to say, oh, I can go back and try tweaking this part of the input and see whether I can get a better result. So that could be useful.''}

Participants also proposed potential ways to improve this local explanation feature. For example: 1) line-level interpretation might be more appropriate than the span level, as P53 in W3-CA said: ``\emph{selecting a span may not be the best option for a user interface; potentially a line is better''}; 2) being able to make immediate changes \textit{''right after identifying the suspicious spots''} via such a tool (P55 in W4-CA); 3) syntax tree driven selection so that \emph{``if you click on the range, you should get the whole function not only the span''} (P63 in W3-CA); 4) adding natural language explanation in addition to the color coding (P71 in W6-NL2Code and P22 in W8-NL2Code).

Moreover, participants saw value in utilizing user edits or interaction with attention distribution as signals to improve the model or future interactions, for example, 
collecting the user edits, as P63 in W3-CA said that \emph{``it might be more useful for how do we improve the model? Cause if somebody... starts typing in, here's the code that I really wanted here. And you can see how does that align and say, wait, why isn't this aligning? It might be more useful is user feedback back to the model builders.''}

\subsection{Social Transparency} 
 
Social transparency (ST) in AI system is recently proposed by \citet{ehsan2021expanding}, built on the concept of social transparency in social computing\cite{stuart2012social}. ST makes visible socio-organizational factors that influence the use of AI by presenting other users' interactions with the AI, aiming to facilitate a better understanding of and effective actions with AI output. To explore ST in the context of GenAI for code, we introduced an open-ended probe in Figure~\ref{fig:social}: an image highlighting that Alex was not working alone but together with a team of software engineers, all using the GenAI. We asked participants to ideate on \textit{what users might want to know about other team members' interactions with the AI to help them better understand and use the AI}. We asked participants to write down their ideas in Mural, and then did the clustering, voting and sharing again. We collected 111 responses about ST from Mural, grouped them to categories, and sorted them based on the frequencies of their occurrences. With the open-ended probe, we were able to elicit user needs for ST in the broad context of software engineer work with the participation of GenAI for code. We therefore map the findings to a stage model of the software development life cycle based on ~\citet{Rani2017ADS}:  1) Requirement Analysis; 2) Design; 3) Implementation; 4) Testing; 5) Deployment and Maintenance~\cite{Rani2017ADS}.
 

At the stage of requirement analysis, participants wanted to know about the \emph{team information} of who, and for what, will be using the AI, such as business requirements and goals of the team, information about each team member (e.g., programming skills and experience), their interaction patterns with the AI (e.g., who is frequently using AI and who is better at using AI) and the reason why they are using the AI. Knowing the team information early on could help users better coordinate and interpret ST information in later stages.

Information about other people's \emph{experience with the AI} could help at the design stage so that users could better design their programming tasks with the GenAI, including taking precautions on when and how they need to take more control on the code generation. It is especially helpful to learn about the AI's performance, capability and limitations from others' experience at this stage. For example, participants wanted to know how other people had evaluated the performance of the model and their judgement of the code quality, including how much time they invested in getting started with AI,  how the AI-generated code compared to human-written code, together with the strengths and weaknesses of the AI. Moreover, they were interested in seeing limitations, such as common pitfalls and errors that other members had experienced. 

 Going into the development stage, participants wanted to have information about \emph{similar tasks or requests} of other people to better understand what the AI model was capable of doing, even re-use the generated content, to help them better manage the programming tasks. Some participants expressed a desire to always have clarity on the \emph{authoring entity} of code (whether from AI, a colleague, or other resources) throughout the development stage to better manage their expectations, actions as well as accountability.

At the testing stage, participants wished to know how other people had tested the AI model as a reference, including what measures they had used in their evaluation of code quality. Many requested to understand AI's \emph{impact on humans} at this stage, i.e. how the introduction of the AI influenced the productivity of the team members, such as team speed and productivity. They also wanted to know whether and how the AI-generated code might be added to the codebase. 


Finally, during the deployment and maintenance stage, participants asked for information about how to customize the model parameters and monitored metrics based on the preferences of the team, and how to collect the team's edits and feedback to improve the model. 

These results move beyond ~\citet{ehsan2021expanding}'s work that defined ST features at the point of AI decision-support. Our results open up the design space to embed rich ST features at different stages of software development. ST features in the context of GenAI for code can not only facilitate users to better understand and use the AI, but also potentially foster a more effective ``human-AI assemblage" where GenAI is introduced as a member or assistant of a software engineer team. We encourage future work to further explore the design space.

\section{Discussion}

\subsection{Informing XAI approaches for GenAI for code}






    
    



The rich and complex inputs and outputs of GenAI, and the tight coupling of the two, make it a prominent need for users to form a clear mental model on the scope, characteristics, and limitations of the input and output spaces, as well as a global view of how the model generates outputs from inputs. 
The observation that participants' explainability needs focused overwhelmingly on input, output and global explanations, suggests a mismatch between what users of GenAI applications need and the technical community's current focus on explanations for representation learning and visualizing representations.

The explainability categories we identified have varied technical feasibility with current techniques, and point to topics that are under-explored for generative AI. For example, for the \emph{Performance} category, existing works have used the Computational Accuracy metric to evaluate generative code models \cite{roziere2020unsupervised, hendrycks2021measuring, austin2021program, chen2021evaluating}, but not other metrics we uncovered regarding the characteristics of the generated artifacts and run-time efficiency.  To understand performance differences and limitations with regard to different types of input, solutions have been explored for natural language generation under Prompt Engineering \cite{liu2021pre, liu2021makes}. Similar studies in the code space are limited to the number of few-shot examples in the prompt, and the effect of length of prompt on performance \cite{austin2021program, chen2021evaluating}. A more detailed analysis of these phenomenon, better grounded in the semantics of programming languages, could be a promising area for future research. For the \emph{Control} category, it is feasible to directly interact with model parameters. However, questions of how to elicit human feedback or edits and incorporate them to train generative models are currently under explored. 

XAI for GenAI needs to be contextualized by the characteristics of the generated artifacts, and the practical and cultural contexts to use these artifacts, i.e. the software engineering domain in our case. For example, participants were interested in understanding the input and output spaces with regard to supported language, frameworks, data structures, and so on. They are also interested in metrics reflecting various aspects of code characteristics and impact on human productivity rather than technical accuracy alone.

Participants' questions reflected their desire for an \textit{actionable} or \textit{utility-oriented} understanding to support their end goal of optimizing code generation and programming productivity \cite{dhanorkar2021needs, liao2021question,liao2021human}, such as asking the \emph{Input, Output, How} and \emph{How-to} XAI questions   (described in Section \ref{sec:general}) 
to facilitate strategies to get better outputs from the AI. This actionable understanding can also be supported by enabling follow-up actions towards their goals after seeing transparent information. For example, participants suggested many interactions to allow them to act on uncertainty information and improve the generated code.

In short, explainability needs should be understood and addressed with the ecological factors in mind, situated in an understanding of the behavioral and social contexts of the system \cite{kogan2020mapping, muller2021designing}. We paid attention to the nature of software engineer tasks and workflows as part of the pragmatic human environment of the system. For example, participants demanded information regarding system requirements and impacts in relation to their broader work environment. The discussions around ST further suggested that users' explainability needs may vary at different stages of software engineering lifecycle, adding \textit{temporal} context to the work contexts. We encourage future research to consider these contextual aspects while designing and evaluating GenAI explanations.

\subsection{Design implications for GenAI for code}






In our study, we observed a positive user reception towards natural language explanations and interactions with the GenAI as user inputs are programming- or natural-language based. We envision that a conversational agent interface could be a natural fit for AI assistant for code or co-programming tools, as explored in recent work~\cite{kuttal2020towards,kuttal2021trade}

We also noticed that the explainability needs and user needs in general may differ for users with varied levels of programming skills. Less experienced participants generally asked fewer questions in the workshop. It is possible that they face more challenges articulating or realizing they had certain explainability needs, and may benefit from more proactive explanations or an entire interaction session focusing on training and explaining rather than in-situ explanations. Future research should further explore user needs in GenAI for code use cases that target reducing programming barriers and enabling novices to code. 


Lastly, the bulk of explainability needs around GenAI for code leads to the question of utility of GenAI for code itself. One concern stems
from the assumption that users are burdened with understanding how the AI works and adjusting their inputs for optimal outcomes, as expressed by P7 in W8-NL2Code: \textit{``as you were talking about this in this session, I keep thinking, if you are learning about all the use cases and how you're going to structure the natural language and everything, how is it kind of different than just learning an actual language?''}. We also note that at the current time, generative code models are prone to errors and will require post-generation improvements from humans~\cite{muller2021datascientists, weisz2021perfection}. There is a fundamental question on the readiness of the technology, which may not be able to 
addressed by providing explainability alone. We urge the AI and HCI community to evaluate 
GenAI for code technologies, define intended use, and refrain from use cases that have not been validated or have risks of harmful consequences for stakeholders. 


\subsection{Human-centered, participatory, and question-driven approaches to XAI design}



Our study used a scenario-based design \cite{rosson2009scenario} combined with a question-driven method to elicit explaiinability needs, based on work by Liao et al.~\cite{liao2020questioning,liao2021question}. We reflect on a few strategies that worked well for our study, and encourage researchers and practitioners to adopt similar approaches to understand users' explainability needs early-on to drive technical and design choices. 

First, we found the use of a scenario, persona and an illustrated prototype corresponding to the scenario to effectively aid the elicitation of rich user questions, even for a technology that was completely new for participants.  As~\citet{liao2021question} suggested ``for highly novel systems, scenarios or low-fi prototypes can be used to elicit questions''. Moreover, we incorporated real outputs from state-of-the-art generative code models in the low-fi prototype to illustrate the model capabilities, which was repeatedly inquired by participants to confirm the scenario has a realistic reflection of the GenAI capabilities. This approach also echos a recent trend in AI design to utilize real data points as 
``data probes'' to aid design ideation~\cite{subramonyam2021towards}. The choice of data point can sway the discussions in the user study. We suggest to choose ones that reflect the AI's true capabilities, including limitations and errors to explore the design space more thoroughly. 

Second, we found the procedure of ``pooling-clustering-voting'' questions to be productive. While ~\citet{liao2021question}'s original method only deals with question elicitation in one-on-one interviews, we adopted a participatory workshop format 
to 
balance between individual brainstorming and group discussions. Giving participants seven minutes to post their questions allowed a quantity of questions to be added. Working collaboratively to cluster them and vote on the clusters encouraged discussions and building upon each other's ideas. This procedure also allowed participants to naturally articulate reasons behind their questions for our data collection.

Lastly, in addition to the question elicitation exercise, we explored four types of XAI features with low-fi prototypes. We defined these features based on prior works in different technical and application domains. A critical study design decision we had to make is the open-endedness of these prototypes. For some features (uncertainty indicators and attention visualizers), there already exist relevant techniques, so we chose a more concrete design to elicit feedback and ideas for extension. For other features (documentation and ST), the design space is less defined so we used more open-ended probes to allow participants to freely and critically brainstorm. In short, the design of a ``probe'' for participatory ideation and feedback is a non-trivial design decision, requiring a balance between concreteness and openness, considerations of both technical feasibility and user value, and diligent pilot testing~\cite{gaver1999probes, boehner2007hci}.

\section{Conclusion}
Despite growing efforts to apply state-of-the-art GenAI models to support software engineering tasks, investigations on user needs for such technologies have been scarce.  Our work is among the first to study users' explainability needs of GenAI for code. By combining scenario-based design~\cite{rosson2009scenario} and a recently proposed question-driven design method for XAI~\cite{liao2020questioning,liao2021question}, we conducted 9 participatory workshops with 43 software engineers to understand their explainability needs with three use cases of GenAI for code: natural language to code, code translation, and code auto-completion. As a result, we identified 11 categories of explainability needs in the context of GenAI for code. We provided detailed definitions and examples for these categories, and contrasted them with common explainability needs for discriminative ML discussed in prior work~\cite{liao2020questioning,lim2010toolkit}. In addition, we proposed four areas of XAI features for these use cases, collected feedback from participants and provided concrete design recommendations. We hope that our results can inspire future AI and HCI work that can enable better human-AI collaboration for software engineering, and encourage more human-centered approaches to drive AI technical development.

\bibliographystyle{ACM-Reference-Format}
\bibliography{reference}


\begin{thebibliography}{105}


\ifx \showCODEN    \undefined \def \showCODEN     #1{\unskip}     \fi
\ifx \showDOI      \undefined \def \showDOI       #1{#1}\fi
\ifx \showISBNx    \undefined \def \showISBNx     #1{\unskip}     \fi
\ifx \showISBNxiii \undefined \def \showISBNxiii  #1{\unskip}     \fi
\ifx \showISSN     \undefined \def \showISSN      #1{\unskip}     \fi
\ifx \showLCCN     \undefined \def \showLCCN      #1{\unskip}     \fi
\ifx \shownote     \undefined \def \shownote      #1{#1}          \fi
\ifx \showarticletitle \undefined \def \showarticletitle #1{#1}   \fi
\ifx \showURL      \undefined \def \showURL       {\relax}        \fi
\providecommand\bibfield[2]{#2}
\providecommand\bibinfo[2]{#2}
\providecommand\natexlab[1]{#1}
\providecommand\showeprint[2][]{arXiv:#2}

\bibitem[\protect\citeauthoryear{Adadi and Berrada}{Adadi and Berrada}{2018}]%
        {adadi2018peeking}
\bibfield{author}{\bibinfo{person}{Amina Adadi} {and} \bibinfo{person}{Mohammed
  Berrada}.} \bibinfo{year}{2018}\natexlab{}.
\newblock \showarticletitle{Peeking inside the black-box: A survey on
  Explainable Artificial Intelligence (XAI)}.
\newblock \bibinfo{journal}{\emph{IEEE Access}}  \bibinfo{volume}{6}
  (\bibinfo{year}{2018}), \bibinfo{pages}{52138--52160}.
\newblock


\bibitem[\protect\citeauthoryear{Agarwal, Talamadupula, Houde, Martinez,
  Muller, Richards, Ross, and Weisz}{Agarwal et~al\mbox{.}}{2020}]%
        {Agarwal2020QualityE}
\bibfield{author}{\bibinfo{person}{Mayank Agarwal}, \bibinfo{person}{Kartik
  Talamadupula}, \bibinfo{person}{Stephanie Houde}, \bibinfo{person}{Fernando
  Martinez}, \bibinfo{person}{Michael~J. Muller}, \bibinfo{person}{John~T.
  Richards}, \bibinfo{person}{Steven Ross}, {and} \bibinfo{person}{Justin~D.
  Weisz}.} \bibinfo{year}{2020}\natexlab{}.
\newblock \showarticletitle{Quality Estimation \& Interpretability for Code
  Translation}.
\newblock \bibinfo{journal}{\emph{ArXiv}}  \bibinfo{volume}{abs/2012.07581}
  (\bibinfo{year}{2020}).
\newblock


\bibitem[\protect\citeauthoryear{Ahmad, Chakraborty, Ray, and Chang}{Ahmad
  et~al\mbox{.}}{2021}]%
        {plbart}
\bibfield{author}{\bibinfo{person}{Wasi Ahmad}, \bibinfo{person}{Saikat
  Chakraborty}, \bibinfo{person}{Baishakhi Ray}, {and} \bibinfo{person}{Kai-Wei
  Chang}.} \bibinfo{year}{2021}\natexlab{}.
\newblock \showarticletitle{Unified Pre-training for Program Understanding and
  Generation}. In \bibinfo{booktitle}{\emph{Proceedings of the 2021 Conference
  of the North American Chapter of the Association for Computational
  Linguistics: Human Language Technologies}}. \bibinfo{publisher}{Association
  for Computational Linguistics}, \bibinfo{address}{Online},
  \bibinfo{pages}{2655--2668}.
\newblock
\urldef\tempurl%
\url{https://doi.org/10.18653/v1/2021.naacl-main.211}
\showDOI{\tempurl}


\bibitem[\protect\citeauthoryear{Allamanis, Barr, Devanbu, and
  Sutton}{Allamanis et~al\mbox{.}}{2018}]%
        {allamanis2018survey}
\bibfield{author}{\bibinfo{person}{Miltiadis Allamanis},
  \bibinfo{person}{Earl~T Barr}, \bibinfo{person}{Premkumar Devanbu}, {and}
  \bibinfo{person}{Charles Sutton}.} \bibinfo{year}{2018}\natexlab{}.
\newblock \showarticletitle{A survey of machine learning for big code and
  naturalness}.
\newblock \bibinfo{journal}{\emph{ACM Computing Surveys (CSUR)}}
  \bibinfo{volume}{51}, \bibinfo{number}{4} (\bibinfo{year}{2018}),
  \bibinfo{pages}{1--37}.
\newblock


\bibitem[\protect\citeauthoryear{Amershi, Cakmak, Knox, and Kulesza}{Amershi
  et~al\mbox{.}}{2014}]%
        {amershi2014power}
\bibfield{author}{\bibinfo{person}{Saleema Amershi}, \bibinfo{person}{Maya
  Cakmak}, \bibinfo{person}{William~Bradley Knox}, {and} \bibinfo{person}{Todd
  Kulesza}.} \bibinfo{year}{2014}\natexlab{}.
\newblock \showarticletitle{Power to the people: The role of humans in
  interactive machine learning}.
\newblock \bibinfo{journal}{\emph{Ai Magazine}} \bibinfo{volume}{35},
  \bibinfo{number}{4} (\bibinfo{year}{2014}), \bibinfo{pages}{105--120}.
\newblock


\bibitem[\protect\citeauthoryear{Aragon, Guha, Kogan, Muller, and Neff}{Aragon
  et~al\mbox{.}}{2022}]%
        {aragon_human-centered_2022}
\bibfield{author}{\bibinfo{person}{Cecilia Aragon}, \bibinfo{person}{Shion
  Guha}, \bibinfo{person}{Marina Kogan}, \bibinfo{person}{Michael Muller},
  {and} \bibinfo{person}{Gina Neff}.} \bibinfo{year}{2022}\natexlab{}.
\newblock \bibinfo{booktitle}{\emph{Human-{Centered} {Data} {Science}: {An}
  {Introduction}}}.
\newblock \bibinfo{publisher}{MIT Press}, \bibinfo{address}{Cambridge, MA}.
\newblock


\bibitem[\protect\citeauthoryear{Aragon, Hutto, Echenique, Fiore-Gartland,
  Huang, Kim, Neff, Xing, and Bayer}{Aragon et~al\mbox{.}}{2016}]%
        {aragon2016developing}
\bibfield{author}{\bibinfo{person}{Cecilia Aragon}, \bibinfo{person}{Clayton
  Hutto}, \bibinfo{person}{Andy Echenique}, \bibinfo{person}{Brittany
  Fiore-Gartland}, \bibinfo{person}{Yun Huang}, \bibinfo{person}{Jinyoung Kim},
  \bibinfo{person}{Gina Neff}, \bibinfo{person}{Wanli Xing}, {and}
  \bibinfo{person}{Joseph Bayer}.} \bibinfo{year}{2016}\natexlab{}.
\newblock \showarticletitle{Developing a research agenda for human-centered
  data science}. In \bibinfo{booktitle}{\emph{Proceedings of the 19th ACM
  Conference on Computer Supported Cooperative Work and Social Computing
  Companion}}. \bibinfo{pages}{529--535}.
\newblock


\bibitem[\protect\citeauthoryear{Arnold, Bellamy, Hind, Houde, Mehta,
  Mojsilovi{\'c}, Nair, Ramamurthy, Olteanu, Piorkowski, et~al\mbox{.}}{Arnold
  et~al\mbox{.}}{2019}]%
        {arnold2019factsheets}
\bibfield{author}{\bibinfo{person}{Matthew Arnold}, \bibinfo{person}{Rachel~KE
  Bellamy}, \bibinfo{person}{Michael Hind}, \bibinfo{person}{Stephanie Houde},
  \bibinfo{person}{Sameep Mehta}, \bibinfo{person}{Aleksandra Mojsilovi{\'c}},
  \bibinfo{person}{Ravi Nair}, \bibinfo{person}{K~Natesan Ramamurthy},
  \bibinfo{person}{Alexandra Olteanu}, \bibinfo{person}{David Piorkowski},
  {et~al\mbox{.}}} \bibinfo{year}{2019}\natexlab{}.
\newblock \showarticletitle{FactSheets: Increasing trust in AI services through
  supplier's declarations of conformity}.
\newblock \bibinfo{journal}{\emph{IBM Journal of Research and Development}}
  \bibinfo{volume}{63}, \bibinfo{number}{4/5} (\bibinfo{year}{2019}),
  \bibinfo{pages}{6--1}.
\newblock


\bibitem[\protect\citeauthoryear{Austin, Odena, Nye, Bosma, Michalewski, Dohan,
  Jiang, Cai, Terry, Le, et~al\mbox{.}}{Austin et~al\mbox{.}}{2021}]%
        {austin2021program}
\bibfield{author}{\bibinfo{person}{Jacob Austin}, \bibinfo{person}{Augustus
  Odena}, \bibinfo{person}{Maxwell Nye}, \bibinfo{person}{Maarten Bosma},
  \bibinfo{person}{Henryk Michalewski}, \bibinfo{person}{David Dohan},
  \bibinfo{person}{Ellen Jiang}, \bibinfo{person}{Carrie Cai},
  \bibinfo{person}{Michael Terry}, \bibinfo{person}{Quoc Le}, {et~al\mbox{.}}}
  \bibinfo{year}{2021}\natexlab{}.
\newblock \showarticletitle{Program Synthesis with Large Language Models}.
\newblock \bibinfo{journal}{\emph{arXiv preprint arXiv:2108.07732}}
  (\bibinfo{year}{2021}).
\newblock


\bibitem[\protect\citeauthoryear{Bhatt, Antor{\'a}n, Zhang, Liao, Sattigeri,
  Fogliato, Melan{\c{c}}on, Krishnan, Stanley, Tickoo, et~al\mbox{.}}{Bhatt
  et~al\mbox{.}}{2021}]%
        {bhatt2021uncertainty}
\bibfield{author}{\bibinfo{person}{Umang Bhatt}, \bibinfo{person}{Javier
  Antor{\'a}n}, \bibinfo{person}{Yunfeng Zhang}, \bibinfo{person}{Q~Vera Liao},
  \bibinfo{person}{Prasanna Sattigeri}, \bibinfo{person}{Riccardo Fogliato},
  \bibinfo{person}{Gabrielle Melan{\c{c}}on}, \bibinfo{person}{Ranganath
  Krishnan}, \bibinfo{person}{Jason Stanley}, \bibinfo{person}{Omesh Tickoo},
  {et~al\mbox{.}}} \bibinfo{year}{2021}\natexlab{}.
\newblock \showarticletitle{Uncertainty as a form of transparency: Measuring,
  communicating, and using uncertainty}. In
  \bibinfo{booktitle}{\emph{Proceedings of the 2021 AAAI/ACM Conference on AI,
  Ethics, and Society}}. \bibinfo{pages}{401--413}.
\newblock


\bibitem[\protect\citeauthoryear{Bhatt, Xiang, Sharma, Weller, Taly, Jia,
  Ghosh, Puri, Moura, and Eckersley}{Bhatt et~al\mbox{.}}{2020}]%
        {bhatt2020explainable}
\bibfield{author}{\bibinfo{person}{Umang Bhatt}, \bibinfo{person}{Alice Xiang},
  \bibinfo{person}{Shubham Sharma}, \bibinfo{person}{Adrian Weller},
  \bibinfo{person}{Ankur Taly}, \bibinfo{person}{Yunhan Jia},
  \bibinfo{person}{Joydeep Ghosh}, \bibinfo{person}{Ruchir Puri},
  \bibinfo{person}{Jos{\'e}~MF Moura}, {and} \bibinfo{person}{Peter
  Eckersley}.} \bibinfo{year}{2020}\natexlab{}.
\newblock \showarticletitle{Explainable machine learning in deployment}. In
  \bibinfo{booktitle}{\emph{Proceedings of the 2020 Conference on Fairness,
  Accountability, and Transparency}}. \bibinfo{pages}{648--657}.
\newblock


\bibitem[\protect\citeauthoryear{Boehner, Vertesi, Sengers, and
  Dourish}{Boehner et~al\mbox{.}}{2007}]%
        {boehner2007hci}
\bibfield{author}{\bibinfo{person}{Kirsten Boehner}, \bibinfo{person}{Janet
  Vertesi}, \bibinfo{person}{Phoebe Sengers}, {and} \bibinfo{person}{Paul
  Dourish}.} \bibinfo{year}{2007}\natexlab{}.
\newblock \showarticletitle{How HCI interprets the probes}. In
  \bibinfo{booktitle}{\emph{Proceedings of the SIGCHI conference on Human
  factors in computing systems}}. \bibinfo{pages}{1077--1086}.
\newblock


\bibitem[\protect\citeauthoryear{Brown, Mann, Ryder, Subbiah, Kaplan, Dhariwal,
  Neelakantan, Shyam, Sastry, Askell, Agarwal, Herbert-Voss, Krueger, Henighan,
  Child, Ramesh, Ziegler, Wu, Winter, Hesse, Chen, Sigler, Litwin, Gray, Chess,
  Clark, Berner, McCandlish, Radford, Sutskever, and Amodei}{Brown
  et~al\mbox{.}}{2020}]%
        {Brown2020LanguageMA}
\bibfield{author}{\bibinfo{person}{Tom~B. Brown}, \bibinfo{person}{Benjamin
  Mann}, \bibinfo{person}{Nick Ryder}, \bibinfo{person}{Melanie Subbiah},
  \bibinfo{person}{Jared Kaplan}, \bibinfo{person}{Prafulla Dhariwal},
  \bibinfo{person}{Arvind Neelakantan}, \bibinfo{person}{Pranav Shyam},
  \bibinfo{person}{Girish Sastry}, \bibinfo{person}{Amanda Askell},
  \bibinfo{person}{Sandhini Agarwal}, \bibinfo{person}{Ariel Herbert-Voss},
  \bibinfo{person}{Gretchen Krueger}, \bibinfo{person}{T.~J. Henighan},
  \bibinfo{person}{Rewon Child}, \bibinfo{person}{Aditya Ramesh},
  \bibinfo{person}{Daniel~M. Ziegler}, \bibinfo{person}{Jeff Wu},
  \bibinfo{person}{Clemens Winter}, \bibinfo{person}{Christopher Hesse},
  \bibinfo{person}{Mark Chen}, \bibinfo{person}{Eric Sigler},
  \bibinfo{person}{Mateusz Litwin}, \bibinfo{person}{Scott Gray},
  \bibinfo{person}{Benjamin Chess}, \bibinfo{person}{Jack Clark},
  \bibinfo{person}{Christopher Berner}, \bibinfo{person}{Sam McCandlish},
  \bibinfo{person}{Alec Radford}, \bibinfo{person}{Ilya Sutskever}, {and}
  \bibinfo{person}{Dario Amodei}.} \bibinfo{year}{2020}\natexlab{}.
\newblock \showarticletitle{Language Models are Few-Shot Learners}.
\newblock \bibinfo{journal}{\emph{ArXiv}}  \bibinfo{volume}{abs/2005.14165}
  (\bibinfo{year}{2020}).
\newblock


\bibitem[\protect\citeauthoryear{Caruana, Lou, Gehrke, Koch, Sturm, and
  Elhadad}{Caruana et~al\mbox{.}}{2015}]%
        {caruana2015intelligible}
\bibfield{author}{\bibinfo{person}{Rich Caruana}, \bibinfo{person}{Yin Lou},
  \bibinfo{person}{Johannes Gehrke}, \bibinfo{person}{Paul Koch},
  \bibinfo{person}{Marc Sturm}, {and} \bibinfo{person}{Noemie Elhadad}.}
  \bibinfo{year}{2015}\natexlab{}.
\newblock \showarticletitle{Intelligible models for healthcare: Predicting
  pneumonia risk and hospital 30-day readmission}. In
  \bibinfo{booktitle}{\emph{Proceedings of KDD}}.
\newblock


\bibitem[\protect\citeauthoryear{Chen, Tworek, Jun, Yuan, de~Oliveira~Pinto,
  Kaplan, Edwards, Burda, Joseph, Brockman, Ray, Puri, Krueger, Petrov, Khlaaf,
  Sastry, Mishkin, Chan, Gray, Ryder, Pavlov, Power, Kaiser, Bavarian, Winter,
  Tillet, Such, Cummings, Plappert, Chantzis, Barnes, Herbert-Voss, Guss,
  Nichol, Paino, Tezak, Tang, Babuschkin, Balaji, Jain, Saunders, Hesse, Carr,
  Leike, Achiam, Misra, Morikawa, Radford, Knight, Brundage, Murati, Mayer,
  Welinder, McGrew, Amodei, McCandlish, Sutskever, and Zaremba}{Chen
  et~al\mbox{.}}{2021}]%
        {chen2021evaluating}
\bibfield{author}{\bibinfo{person}{Mark Chen}, \bibinfo{person}{Jerry Tworek},
  \bibinfo{person}{Heewoo Jun}, \bibinfo{person}{Qiming Yuan},
  \bibinfo{person}{Henrique~Ponde de Oliveira~Pinto}, \bibinfo{person}{Jared
  Kaplan}, \bibinfo{person}{Harri Edwards}, \bibinfo{person}{Yuri Burda},
  \bibinfo{person}{Nicholas Joseph}, \bibinfo{person}{Greg Brockman},
  \bibinfo{person}{Alex Ray}, \bibinfo{person}{Raul Puri},
  \bibinfo{person}{Gretchen Krueger}, \bibinfo{person}{Michael Petrov},
  \bibinfo{person}{Heidy Khlaaf}, \bibinfo{person}{Girish Sastry},
  \bibinfo{person}{Pamela Mishkin}, \bibinfo{person}{Brooke Chan},
  \bibinfo{person}{Scott Gray}, \bibinfo{person}{Nick Ryder},
  \bibinfo{person}{Mikhail Pavlov}, \bibinfo{person}{Alethea Power},
  \bibinfo{person}{Lukasz Kaiser}, \bibinfo{person}{Mohammad Bavarian},
  \bibinfo{person}{Clemens Winter}, \bibinfo{person}{Philippe Tillet},
  \bibinfo{person}{Felipe~Petroski Such}, \bibinfo{person}{Dave Cummings},
  \bibinfo{person}{Matthias Plappert}, \bibinfo{person}{Fotios Chantzis},
  \bibinfo{person}{Elizabeth Barnes}, \bibinfo{person}{Ariel Herbert-Voss},
  \bibinfo{person}{William~Hebgen Guss}, \bibinfo{person}{Alex Nichol},
  \bibinfo{person}{Alex Paino}, \bibinfo{person}{Nikolas Tezak},
  \bibinfo{person}{Jie Tang}, \bibinfo{person}{Igor Babuschkin},
  \bibinfo{person}{Suchir Balaji}, \bibinfo{person}{Shantanu Jain},
  \bibinfo{person}{William Saunders}, \bibinfo{person}{Christopher Hesse},
  \bibinfo{person}{Andrew~N. Carr}, \bibinfo{person}{Jan Leike},
  \bibinfo{person}{Josh Achiam}, \bibinfo{person}{Vedant Misra},
  \bibinfo{person}{Evan Morikawa}, \bibinfo{person}{Alec Radford},
  \bibinfo{person}{Matthew Knight}, \bibinfo{person}{Miles Brundage},
  \bibinfo{person}{Mira Murati}, \bibinfo{person}{Katie Mayer},
  \bibinfo{person}{Peter Welinder}, \bibinfo{person}{Bob McGrew},
  \bibinfo{person}{Dario Amodei}, \bibinfo{person}{Sam McCandlish},
  \bibinfo{person}{Ilya Sutskever}, {and} \bibinfo{person}{Wojciech Zaremba}.}
  \bibinfo{year}{2021}\natexlab{}.
\newblock \bibinfo{title}{Evaluating Large Language Models Trained on Code}.
\newblock
\newblock
\showeprint[arxiv]{2107.03374}~[cs.LG]


\bibitem[\protect\citeauthoryear{Chen, Li, Grosse, and Duvenaud}{Chen
  et~al\mbox{.}}{2018}]%
        {Chen2018IsolatingSO}
\bibfield{author}{\bibinfo{person}{Tian~Qi Chen}, \bibinfo{person}{Xuechen Li},
  \bibinfo{person}{Roger~B. Grosse}, {and} \bibinfo{person}{David~Kristjanson
  Duvenaud}.} \bibinfo{year}{2018}\natexlab{}.
\newblock \showarticletitle{Isolating Sources of Disentanglement in Variational
  Autoencoders}. In \bibinfo{booktitle}{\emph{NeurIPS}}.
\newblock


\bibitem[\protect\citeauthoryear{Devanbu}{Devanbu}{2015}]%
        {devanbu2015new}
\bibfield{author}{\bibinfo{person}{Premkumar Devanbu}.}
  \bibinfo{year}{2015}\natexlab{}.
\newblock \showarticletitle{New initiative: The naturalness of software}. In
  \bibinfo{booktitle}{\emph{{2015 IEEE/ACM 37th IEEE International Conference
  on Software Engineering}}}, Vol.~\bibinfo{volume}{2}. IEEE,
  \bibinfo{pages}{543--546}.
\newblock


\bibitem[\protect\citeauthoryear{Dhanorkar, Wolf, Qian, Xu, Popa, and
  Li}{Dhanorkar et~al\mbox{.}}{2021}]%
        {dhanorkar2021needs}
\bibfield{author}{\bibinfo{person}{Shipi Dhanorkar},
  \bibinfo{person}{Christine~T Wolf}, \bibinfo{person}{Kun Qian},
  \bibinfo{person}{Anbang Xu}, \bibinfo{person}{Lucian Popa}, {and}
  \bibinfo{person}{Yunyao Li}.} \bibinfo{year}{2021}\natexlab{}.
\newblock \showarticletitle{Who needs to know what, when?: Broadening the
  Explainable AI (XAI) Design Space by Looking at Explanations Across the AI
  Lifecycle}. In \bibinfo{booktitle}{\emph{Designing Interactive Systems
  Conference 2021}}. \bibinfo{pages}{1591--1602}.
\newblock


\bibitem[\protect\citeauthoryear{Ehsan, Liao, Muller, Riedl, and Weisz}{Ehsan
  et~al\mbox{.}}{2021a}]%
        {ehsan2021expanding}
\bibfield{author}{\bibinfo{person}{Upol Ehsan}, \bibinfo{person}{Q~Vera Liao},
  \bibinfo{person}{Michael Muller}, \bibinfo{person}{Mark~O Riedl}, {and}
  \bibinfo{person}{Justin~D Weisz}.} \bibinfo{year}{2021}\natexlab{a}.
\newblock \showarticletitle{Expanding explainability: Towards social
  transparency in ai systems}. In \bibinfo{booktitle}{\emph{Proceedings of the
  2021 CHI Conference on Human Factors in Computing Systems}}.
  \bibinfo{pages}{1--19}.
\newblock


\bibitem[\protect\citeauthoryear{Ehsan and Riedl}{Ehsan and Riedl}{2021}]%
        {ehsan2021pitfalls}
\bibfield{author}{\bibinfo{person}{Upol Ehsan} {and} \bibinfo{person}{Mark
  Riedl}.} \bibinfo{year}{2021}\natexlab{}.
\newblock \bibinfo{booktitle}{\emph{Explainability Pitfalls: Beyond Dark
  Patterns in Explainable AI - paper at HCAI@NeurIPS2021 workshop on human
  centered AI}}.
\newblock
\urldef\tempurl%
\url{https://sites.google.com/view/hcai-human-centered-ai-neurips/home}
\showURL{%
\tempurl}
\newblock
\shownote{Accessed January 19, 2022.}


\bibitem[\protect\citeauthoryear{Ehsan and Riedl}{Ehsan and Riedl}{2020}]%
        {ehsan2020human}
\bibfield{author}{\bibinfo{person}{Upol Ehsan} {and} \bibinfo{person}{Mark~O
  Riedl}.} \bibinfo{year}{2020}\natexlab{}.
\newblock \showarticletitle{Human-centered explainable ai: Towards a reflective
  sociotechnical approach}. In \bibinfo{booktitle}{\emph{International
  Conference on Human-Computer Interaction}}. Springer,
  \bibinfo{pages}{449--466}.
\newblock


\bibitem[\protect\citeauthoryear{Ehsan, Wintersberger, Liao, Mara, Streit,
  Wachter, Riener, and Riedl}{Ehsan et~al\mbox{.}}{2021b}]%
        {ehsan2021operationalizing}
\bibfield{author}{\bibinfo{person}{Upol Ehsan}, \bibinfo{person}{Philipp
  Wintersberger}, \bibinfo{person}{Q~Vera Liao}, \bibinfo{person}{Martina
  Mara}, \bibinfo{person}{Marc Streit}, \bibinfo{person}{Sandra Wachter},
  \bibinfo{person}{Andreas Riener}, {and} \bibinfo{person}{Mark~O Riedl}.}
  \bibinfo{year}{2021}\natexlab{b}.
\newblock \showarticletitle{Operationalizing Human-Centered Perspectives in
  Explainable AI}. In \bibinfo{booktitle}{\emph{Extended Abstracts of the 2021
  CHI Conference on Human Factors in Computing Systems}}.
  \bibinfo{pages}{1--6}.
\newblock


\bibitem[\protect\citeauthoryear{Feng, Guo, Tang, Duan, Feng, Gong, Shou, Qin,
  Liu, Jiang, et~al\mbox{.}}{Feng et~al\mbox{.}}{2020}]%
        {feng2020codebert}
\bibfield{author}{\bibinfo{person}{Zhangyin Feng}, \bibinfo{person}{Daya Guo},
  \bibinfo{person}{Duyu Tang}, \bibinfo{person}{Nan Duan},
  \bibinfo{person}{Xiaocheng Feng}, \bibinfo{person}{Ming Gong},
  \bibinfo{person}{Linjun Shou}, \bibinfo{person}{Bing Qin},
  \bibinfo{person}{Ting Liu}, \bibinfo{person}{Daxin Jiang}, {et~al\mbox{.}}}
  \bibinfo{year}{2020}\natexlab{}.
\newblock \showarticletitle{Codebert: A pre-trained model for programming and
  natural languages}.
\newblock \bibinfo{journal}{\emph{arXiv preprint arXiv:2002.08155}}
  (\bibinfo{year}{2020}).
\newblock


\bibitem[\protect\citeauthoryear{Gaver, Dunne, and Pacenti}{Gaver
  et~al\mbox{.}}{1999}]%
        {gaver1999probes}
\bibfield{author}{\bibinfo{person}{Bill Gaver}, \bibinfo{person}{Tony Dunne},
  {and} \bibinfo{person}{Elena Pacenti}.} \bibinfo{year}{1999}\natexlab{}.
\newblock \showarticletitle{Design: Cultural Probes}.
\newblock \bibinfo{journal}{\emph{Interactions}} \bibinfo{volume}{6},
  \bibinfo{number}{1} (\bibinfo{date}{jan} \bibinfo{year}{1999}),
  \bibinfo{pages}{21–29}.
\newblock
\showISSN{1072-5520}
\urldef\tempurl%
\url{https://doi.org/10.1145/291224.291235}
\showDOI{\tempurl}


\bibitem[\protect\citeauthoryear{Geyer, Chilton, Weisz, and Maher}{Geyer
  et~al\mbox{.}}{2021}]%
        {geyer2021hai}
\bibfield{author}{\bibinfo{person}{Werner Geyer}, \bibinfo{person}{Lydia~B
  Chilton}, \bibinfo{person}{Justin~D Weisz}, {and} \bibinfo{person}{Mary~Lou
  Maher}.} \bibinfo{year}{2021}\natexlab{}.
\newblock \showarticletitle{HAI-GEN 2021: 2nd Workshop on Human-AI Co-Creation
  with Generative Models}. In \bibinfo{booktitle}{\emph{26th International
  Conference on Intelligent User Interfaces}}. \bibinfo{pages}{15--17}.
\newblock


\bibitem[\protect\citeauthoryear{Ghosh, Liao, Ramamurthy, Navratil, Sattigeri,
  Varshney, and Zhang}{Ghosh et~al\mbox{.}}{2021}]%
        {ghosh2021uncertainty}
\bibfield{author}{\bibinfo{person}{Soumya Ghosh}, \bibinfo{person}{Q~Vera
  Liao}, \bibinfo{person}{Karthikeyan~Natesan Ramamurthy},
  \bibinfo{person}{Jiri Navratil}, \bibinfo{person}{Prasanna Sattigeri},
  \bibinfo{person}{Kush~R Varshney}, {and} \bibinfo{person}{Yunfeng Zhang}.}
  \bibinfo{year}{2021}\natexlab{}.
\newblock \showarticletitle{Uncertainty Quantification 360: A Holistic Toolkit
  for Quantifying and Communicating the Uncertainty of AI}.
\newblock \bibinfo{journal}{\emph{arXiv preprint arXiv:2106.01410}}
  (\bibinfo{year}{2021}).
\newblock


\bibitem[\protect\citeauthoryear{Github}{Github}{2021}]%
        {web:copilot}
\bibfield{author}{\bibinfo{person}{Github}.} \bibinfo{year}{2021}\natexlab{}.
\newblock \bibinfo{title}{Copilot}.
\newblock
\newblock
\urldef\tempurl%
\url{https://copilot.github.com}
\showURL{%
Retrieved 03-August-2021 from \tempurl}


\bibitem[\protect\citeauthoryear{Goodfellow, Pouget-Abadie, Mirza, Xu,
  Warde-Farley, Ozair, Courville, and Bengio}{Goodfellow et~al\mbox{.}}{2020}]%
        {goodfellow2020generative}
\bibfield{author}{\bibinfo{person}{Ian Goodfellow}, \bibinfo{person}{Jean
  Pouget-Abadie}, \bibinfo{person}{Mehdi Mirza}, \bibinfo{person}{Bing Xu},
  \bibinfo{person}{David Warde-Farley}, \bibinfo{person}{Sherjil Ozair},
  \bibinfo{person}{Aaron Courville}, {and} \bibinfo{person}{Yoshua Bengio}.}
  \bibinfo{year}{2020}\natexlab{}.
\newblock \showarticletitle{Generative adversarial networks}.
\newblock \bibinfo{journal}{\emph{Commun. ACM}} \bibinfo{volume}{63},
  \bibinfo{number}{11} (\bibinfo{year}{2020}), \bibinfo{pages}{139--144}.
\newblock


\bibitem[\protect\citeauthoryear{Guidotti, Monreale, Ruggieri, Turini,
  Giannotti, and Pedreschi}{Guidotti et~al\mbox{.}}{2018}]%
        {guidotti2018survey}
\bibfield{author}{\bibinfo{person}{Riccardo Guidotti}, \bibinfo{person}{Anna
  Monreale}, \bibinfo{person}{Salvatore Ruggieri}, \bibinfo{person}{Franco
  Turini}, \bibinfo{person}{Fosca Giannotti}, {and} \bibinfo{person}{Dino
  Pedreschi}.} \bibinfo{year}{2018}\natexlab{}.
\newblock \showarticletitle{A survey of methods for explaining black box
  models}.
\newblock \bibinfo{journal}{\emph{ACM computing surveys (CSUR)}}
  \bibinfo{volume}{51}, \bibinfo{number}{5} (\bibinfo{year}{2018}),
  \bibinfo{pages}{1--42}.
\newblock


\bibitem[\protect\citeauthoryear{Guo, Ren, Lu, Feng, Tang, Liu, Zhou, Duan,
  Svyatkovskiy, Fu, et~al\mbox{.}}{Guo et~al\mbox{.}}{2020}]%
        {guo2020graphcodebert}
\bibfield{author}{\bibinfo{person}{Daya Guo}, \bibinfo{person}{Shuo Ren},
  \bibinfo{person}{Shuai Lu}, \bibinfo{person}{Zhangyin Feng},
  \bibinfo{person}{Duyu Tang}, \bibinfo{person}{Shujie Liu},
  \bibinfo{person}{Long Zhou}, \bibinfo{person}{Nan Duan},
  \bibinfo{person}{Alexey Svyatkovskiy}, \bibinfo{person}{Shengyu Fu},
  {et~al\mbox{.}}} \bibinfo{year}{2020}\natexlab{}.
\newblock \showarticletitle{Graphcodebert: Pre-training code representations
  with data flow}.
\newblock \bibinfo{journal}{\emph{arXiv preprint arXiv:2009.08366}}
  (\bibinfo{year}{2020}).
\newblock


\bibitem[\protect\citeauthoryear{Guo, Du, Malik, Koh, Kim, Liu, Kim, Zha, and
  Cao}{Guo et~al\mbox{.}}{2019}]%
        {guo2019visualizing}
\bibfield{author}{\bibinfo{person}{Shunan Guo}, \bibinfo{person}{Fan Du},
  \bibinfo{person}{Sana Malik}, \bibinfo{person}{Eunyee Koh},
  \bibinfo{person}{Sungchul Kim}, \bibinfo{person}{Zhicheng Liu},
  \bibinfo{person}{Donghyun Kim}, \bibinfo{person}{Hongyuan Zha}, {and}
  \bibinfo{person}{Nan Cao}.} \bibinfo{year}{2019}\natexlab{}.
\newblock \showarticletitle{Visualizing uncertainty and alternatives in event
  sequence predictions}. In \bibinfo{booktitle}{\emph{Proceedings of the 2019
  CHI Conference on Human Factors in Computing Systems}}.
  \bibinfo{pages}{1--12}.
\newblock


\bibitem[\protect\citeauthoryear{Halfaker and Geiger}{Halfaker and
  Geiger}{2020}]%
        {halfaker2020ores}
\bibfield{author}{\bibinfo{person}{Aaron Halfaker} {and}
  \bibinfo{person}{R~Stuart Geiger}.} \bibinfo{year}{2020}\natexlab{}.
\newblock \showarticletitle{Ores: Lowering barriers with participatory machine
  learning in wikipedia}.
\newblock \bibinfo{journal}{\emph{Proceedings of the ACM on Human-Computer
  Interaction}} \bibinfo{volume}{4}, \bibinfo{number}{CSCW2}
  (\bibinfo{year}{2020}), \bibinfo{pages}{1--37}.
\newblock


\bibitem[\protect\citeauthoryear{Hendrycks, Basart, Kadavath, Mazeika, Arora,
  Guo, Burns, Puranik, He, Song, et~al\mbox{.}}{Hendrycks
  et~al\mbox{.}}{2021}]%
        {hendrycks2021measuring}
\bibfield{author}{\bibinfo{person}{Dan Hendrycks}, \bibinfo{person}{Steven
  Basart}, \bibinfo{person}{Saurav Kadavath}, \bibinfo{person}{Mantas Mazeika},
  \bibinfo{person}{Akul Arora}, \bibinfo{person}{Ethan Guo},
  \bibinfo{person}{Collin Burns}, \bibinfo{person}{Samir Puranik},
  \bibinfo{person}{Horace He}, \bibinfo{person}{Dawn Song}, {et~al\mbox{.}}}
  \bibinfo{year}{2021}\natexlab{}.
\newblock \showarticletitle{Measuring Coding Challenge Competence With APPS}.
\newblock \bibinfo{journal}{\emph{arXiv preprint arXiv:2105.09938}}
  (\bibinfo{year}{2021}).
\newblock


\bibitem[\protect\citeauthoryear{Hilton}{Hilton}{1990}]%
        {hilton1990conversational}
\bibfield{author}{\bibinfo{person}{Denis~J Hilton}.}
  \bibinfo{year}{1990}\natexlab{}.
\newblock \showarticletitle{Conversational processes and causal explanation.}
\newblock \bibinfo{journal}{\emph{Psychological Bulletin}}
  \bibinfo{volume}{107}, \bibinfo{number}{1} (\bibinfo{year}{1990}),
  \bibinfo{pages}{65}.
\newblock


\bibitem[\protect\citeauthoryear{Hind, Houde, Martino, Mojsilovic, Piorkowski,
  Richards, and Varshney}{Hind et~al\mbox{.}}{2020}]%
        {Hind2020ExperiencesWI}
\bibfield{author}{\bibinfo{person}{M. Hind}, \bibinfo{person}{Stephanie Houde},
  \bibinfo{person}{Jacquelyn Martino}, \bibinfo{person}{A. Mojsilovic},
  \bibinfo{person}{David Piorkowski}, \bibinfo{person}{John~T. Richards}, {and}
  \bibinfo{person}{K. Varshney}.} \bibinfo{year}{2020}\natexlab{}.
\newblock \showarticletitle{Experiences with Improving the Transparency of AI
  Models and Services}.
\newblock \bibinfo{journal}{\emph{Extended Abstracts of the 2020 CHI Conference
  on Human Factors in Computing Systems}} (\bibinfo{year}{2020}).
\newblock


\bibitem[\protect\citeauthoryear{Hind, Mehta, Mojsilovic, Nair, Ramamurthy,
  Olteanu, and Varshney}{Hind et~al\mbox{.}}{2019}]%
        {Hind2019IncreasingTI}
\bibfield{author}{\bibinfo{person}{M. Hind}, \bibinfo{person}{S. Mehta},
  \bibinfo{person}{A. Mojsilovic}, \bibinfo{person}{R. Nair},
  \bibinfo{person}{K. Ramamurthy}, \bibinfo{person}{Alexandra Olteanu}, {and}
  \bibinfo{person}{K. Varshney}.} \bibinfo{year}{2019}\natexlab{}.
\newblock \showarticletitle{Increasing Trust in AI Services through Supplier's
  Declarations of Conformity}.
\newblock \bibinfo{journal}{\emph{IBM J. Res. Dev.}}  \bibinfo{volume}{63}
  (\bibinfo{year}{2019}), \bibinfo{pages}{6:1--6:13}.
\newblock


\bibitem[\protect\citeauthoryear{Hindle, Barr, Gabel, Su, and Devanbu}{Hindle
  et~al\mbox{.}}{2016}]%
        {hindle2016naturalness}
\bibfield{author}{\bibinfo{person}{Abram Hindle}, \bibinfo{person}{Earl~T
  Barr}, \bibinfo{person}{Mark Gabel}, \bibinfo{person}{Zhendong Su}, {and}
  \bibinfo{person}{Premkumar Devanbu}.} \bibinfo{year}{2016}\natexlab{}.
\newblock \showarticletitle{On the naturalness of software}.
\newblock \bibinfo{journal}{\emph{Commun. ACM}} \bibinfo{volume}{59},
  \bibinfo{number}{5} (\bibinfo{year}{2016}), \bibinfo{pages}{122--131}.
\newblock


\bibitem[\protect\citeauthoryear{Horvitz}{Horvitz}{1999}]%
        {horvitz1999principles}
\bibfield{author}{\bibinfo{person}{Eric Horvitz}.}
  \bibinfo{year}{1999}\natexlab{}.
\newblock \showarticletitle{Principles of Mixed-Initiative User Interfaces}. In
  \bibinfo{booktitle}{\emph{Proceedings of the SIGCHI Conference on Human
  Factors in Computing Systems}} (Pittsburgh, Pennsylvania, USA)
  \emph{(\bibinfo{series}{CHI '99})}. \bibinfo{publisher}{Association for
  Computing Machinery}, \bibinfo{address}{New York, NY, USA},
  \bibinfo{pages}{159–166}.
\newblock
\showISBNx{0201485591}
\urldef\tempurl%
\url{https://doi.org/10.1145/302979.303030}
\showDOI{\tempurl}


\bibitem[\protect\citeauthoryear{Kim, Zhao, Tian, and Chandra}{Kim
  et~al\mbox{.}}{2021}]%
        {kim2021code}
\bibfield{author}{\bibinfo{person}{Seohyun Kim}, \bibinfo{person}{Jinman Zhao},
  \bibinfo{person}{Yuchi Tian}, {and} \bibinfo{person}{Satish Chandra}.}
  \bibinfo{year}{2021}\natexlab{}.
\newblock \showarticletitle{Code prediction by feeding trees to transformers}.
  In \bibinfo{booktitle}{\emph{2021 IEEE/ACM 43rd International Conference on
  Software Engineering (ICSE)}}. IEEE, \bibinfo{pages}{150--162}.
\newblock


\bibitem[\protect\citeauthoryear{Knowles and Richards}{Knowles and
  Richards}{2021}]%
        {Knowles2021TheSO}
\bibfield{author}{\bibinfo{person}{Bran Knowles} {and} \bibinfo{person}{John~T.
  Richards}.} \bibinfo{year}{2021}\natexlab{}.
\newblock \showarticletitle{The Sanction of Authority: Promoting Public Trust
  in AI}.
\newblock \bibinfo{journal}{\emph{Proceedings of the 2021 ACM Conference on
  Fairness, Accountability, and Transparency}} (\bibinfo{year}{2021}).
\newblock


\bibitem[\protect\citeauthoryear{Kogan, Halfaker, Guha, Aragon, Muller, and
  Geiger}{Kogan et~al\mbox{.}}{2020}]%
        {kogan2020mapping}
\bibfield{author}{\bibinfo{person}{Marina Kogan}, \bibinfo{person}{Aaron
  Halfaker}, \bibinfo{person}{Shion Guha}, \bibinfo{person}{Cecilia Aragon},
  \bibinfo{person}{Michael Muller}, {and} \bibinfo{person}{Stuart Geiger}.}
  \bibinfo{year}{2020}\natexlab{}.
\newblock \showarticletitle{Mapping Out Human-Centered Data Science: Methods,
  Approaches, and Best Practices}. In \bibinfo{booktitle}{\emph{Companion of
  the 2020 ACM International Conference on Supporting Group Work}}.
  \bibinfo{pages}{151--156}.
\newblock


\bibitem[\protect\citeauthoryear{Kuttal, Myers, Gurka, Magar, Piorkowski, and
  Bellamy}{Kuttal et~al\mbox{.}}{2020}]%
        {kuttal2020towards}
\bibfield{author}{\bibinfo{person}{Sandeep~Kaur Kuttal}, \bibinfo{person}{Jarow
  Myers}, \bibinfo{person}{Sam Gurka}, \bibinfo{person}{David Magar},
  \bibinfo{person}{David Piorkowski}, {and} \bibinfo{person}{Rachel Bellamy}.}
  \bibinfo{year}{2020}\natexlab{}.
\newblock \showarticletitle{Towards designing conversational agents for pair
  programming: Accounting for creativity strategies and conversational styles}.
  In \bibinfo{booktitle}{\emph{2020 IEEE Symposium on Visual Languages and
  Human-Centric Computing (VL/HCC)}}. IEEE, \bibinfo{pages}{1--11}.
\newblock


\bibitem[\protect\citeauthoryear{Kuttal, Ong, Kwasny, and Robe}{Kuttal
  et~al\mbox{.}}{2021}]%
        {kuttal2021trade}
\bibfield{author}{\bibinfo{person}{Sandeep~Kaur Kuttal}, \bibinfo{person}{Bali
  Ong}, \bibinfo{person}{Kate Kwasny}, {and} \bibinfo{person}{Peter Robe}.}
  \bibinfo{year}{2021}\natexlab{}.
\newblock \showarticletitle{Trade-offs for Substituting a Human with an Agent
  in a Pair Programming Context: The Good, the Bad, and the Ugly}. In
  \bibinfo{booktitle}{\emph{Proceedings of the 2021 CHI Conference on Human
  Factors in Computing Systems}}. \bibinfo{pages}{1--20}.
\newblock


\bibitem[\protect\citeauthoryear{Lakkaraju, Bach, and Leskovec}{Lakkaraju
  et~al\mbox{.}}{2016}]%
        {lakkaraju2016interpretable}
\bibfield{author}{\bibinfo{person}{Himabindu Lakkaraju},
  \bibinfo{person}{Stephen~H. Bach}, {and} \bibinfo{person}{Jure Leskovec}.}
  \bibinfo{year}{2016}\natexlab{}.
\newblock \showarticletitle{Interpretable Decision Sets: A Joint Framework for
  Description and Prediction}. In \bibinfo{booktitle}{\emph{Proceedings of the
  22nd ACM SIGKDD International Conference on Knowledge Discovery and Data
  Mining}} (San Francisco, California, USA) \emph{(\bibinfo{series}{KDD '16})}.
  \bibinfo{publisher}{Association for Computing Machinery},
  \bibinfo{address}{New York, NY, USA}, \bibinfo{pages}{1675–1684}.
\newblock
\showISBNx{9781450342322}
\urldef\tempurl%
\url{https://doi.org/10.1145/2939672.2939874}
\showDOI{\tempurl}


\bibitem[\protect\citeauthoryear{Lee, Grgi{\'c}-Hla{\v{c}}a, Tschantz, Binns,
  Weller, Carney, and Inkpen}{Lee et~al\mbox{.}}{2020}]%
        {lee2020human}
\bibfield{author}{\bibinfo{person}{Min~Kyung Lee}, \bibinfo{person}{Nina
  Grgi{\'c}-Hla{\v{c}}a}, \bibinfo{person}{Michael~Carl Tschantz},
  \bibinfo{person}{Reuben Binns}, \bibinfo{person}{Adrian Weller},
  \bibinfo{person}{Michelle Carney}, {and} \bibinfo{person}{Kori Inkpen}.}
  \bibinfo{year}{2020}\natexlab{}.
\newblock \showarticletitle{Human-centered approaches to fair and responsible
  AI}. In \bibinfo{booktitle}{\emph{Extended Abstracts of the 2020 CHI
  Conference on Human Factors in Computing Systems}}. \bibinfo{pages}{1--8}.
\newblock


\bibitem[\protect\citeauthoryear{Lee, Kusbit, Kahng, Kim, Yuan, Chan, See,
  Noothigattu, Lee, Psomas, et~al\mbox{.}}{Lee et~al\mbox{.}}{2019}]%
        {lee2019webuildai}
\bibfield{author}{\bibinfo{person}{Min~Kyung Lee}, \bibinfo{person}{Daniel
  Kusbit}, \bibinfo{person}{Anson Kahng}, \bibinfo{person}{Ji~Tae Kim},
  \bibinfo{person}{Xinran Yuan}, \bibinfo{person}{Allissa Chan},
  \bibinfo{person}{Daniel See}, \bibinfo{person}{Ritesh Noothigattu},
  \bibinfo{person}{Siheon Lee}, \bibinfo{person}{Alexandros Psomas},
  {et~al\mbox{.}}} \bibinfo{year}{2019}\natexlab{}.
\newblock \showarticletitle{WeBuildAI: Participatory framework for algorithmic
  governance}.
\newblock \bibinfo{journal}{\emph{Proceedings of the ACM on Human-Computer
  Interaction}} \bibinfo{volume}{3}, \bibinfo{number}{CSCW}
  (\bibinfo{year}{2019}), \bibinfo{pages}{1--35}.
\newblock


\bibitem[\protect\citeauthoryear{Lei, Barzilay, and Jaakkola}{Lei
  et~al\mbox{.}}{2016}]%
        {Lei2016RationalizingNP}
\bibfield{author}{\bibinfo{person}{Tao Lei}, \bibinfo{person}{Regina Barzilay},
  {and} \bibinfo{person}{T. Jaakkola}.} \bibinfo{year}{2016}\natexlab{}.
\newblock \showarticletitle{Rationalizing Neural Predictions}. In
  \bibinfo{booktitle}{\emph{EMNLP}}.
\newblock


\bibitem[\protect\citeauthoryear{Liao, Gruen, and Miller}{Liao
  et~al\mbox{.}}{2020}]%
        {liao2020questioning}
\bibfield{author}{\bibinfo{person}{Q~Vera Liao}, \bibinfo{person}{Daniel
  Gruen}, {and} \bibinfo{person}{Sarah Miller}.}
  \bibinfo{year}{2020}\natexlab{}.
\newblock \showarticletitle{Questioning the AI: informing design practices for
  explainable AI user experiences}. In \bibinfo{booktitle}{\emph{Proceedings of
  the 2020 CHI Conference on Human Factors in Computing Systems}}.
  \bibinfo{pages}{1--15}.
\newblock


\bibitem[\protect\citeauthoryear{Liao and Muller}{Liao and Muller}{2019}]%
        {liao2019enabling}
\bibfield{author}{\bibinfo{person}{Q~Vera Liao} {and} \bibinfo{person}{Michael
  Muller}.} \bibinfo{year}{2019}\natexlab{}.
\newblock \showarticletitle{Enabling Value Sensitive AI Systems through
  Participatory Design Fictions}.
\newblock \bibinfo{journal}{\emph{arXiv preprint arXiv:1912.07381}}
  (\bibinfo{year}{2019}).
\newblock


\bibitem[\protect\citeauthoryear{Liao, Pribi{\'c}, Han, Miller, and Sow}{Liao
  et~al\mbox{.}}{2021a}]%
        {liao2021question}
\bibfield{author}{\bibinfo{person}{Q~Vera Liao}, \bibinfo{person}{Milena
  Pribi{\'c}}, \bibinfo{person}{Jaesik Han}, \bibinfo{person}{Sarah Miller},
  {and} \bibinfo{person}{Daby Sow}.} \bibinfo{year}{2021}\natexlab{a}.
\newblock \showarticletitle{Question-Driven Design Process for Explainable AI
  User Experiences}.
\newblock \bibinfo{journal}{\emph{arXiv preprint arXiv:2104.03483}}
  (\bibinfo{year}{2021}).
\newblock


\bibitem[\protect\citeauthoryear{Liao, Singh, Zhang, and Bellamy}{Liao
  et~al\mbox{.}}{2021b}]%
        {liao2021introduction}
\bibfield{author}{\bibinfo{person}{Q~Vera Liao}, \bibinfo{person}{Moninder
  Singh}, \bibinfo{person}{Yunfeng Zhang}, {and} \bibinfo{person}{Rachel
  Bellamy}.} \bibinfo{year}{2021}\natexlab{b}.
\newblock \showarticletitle{Introduction to explainable ai}. In
  \bibinfo{booktitle}{\emph{Extended Abstracts of the 2021 CHI Conference on
  Human Factors in Computing Systems}}. \bibinfo{pages}{1--3}.
\newblock


\bibitem[\protect\citeauthoryear{Liao and Varshney}{Liao and Varshney}{2021}]%
        {liao2021human}
\bibfield{author}{\bibinfo{person}{Q~Vera Liao} {and} \bibinfo{person}{Kush~R
  Varshney}.} \bibinfo{year}{2021}\natexlab{}.
\newblock \showarticletitle{Human-Centered Explainable AI (XAI): From
  Algorithms to User Experiences}.
\newblock \bibinfo{journal}{\emph{arXiv preprint arXiv:2110.10790}}
  (\bibinfo{year}{2021}).
\newblock


\bibitem[\protect\citeauthoryear{Lim and Dey}{Lim and Dey}{2010}]%
        {lim2010toolkit}
\bibfield{author}{\bibinfo{person}{Brian~Y Lim} {and} \bibinfo{person}{Anind~K
  Dey}.} \bibinfo{year}{2010}\natexlab{}.
\newblock \showarticletitle{Toolkit to support intelligibility in context-aware
  applications}. In \bibinfo{booktitle}{\emph{Proceedings of the 12th ACM
  international conference on Ubiquitous computing}}. \bibinfo{pages}{13--22}.
\newblock


\bibitem[\protect\citeauthoryear{Lim, Dey, and Avrahami}{Lim
  et~al\mbox{.}}{2009}]%
        {lim2009and}
\bibfield{author}{\bibinfo{person}{Brian~Y Lim}, \bibinfo{person}{Anind~K Dey},
  {and} \bibinfo{person}{Daniel Avrahami}.} \bibinfo{year}{2009}\natexlab{}.
\newblock \showarticletitle{Why and why not explanations improve the
  intelligibility of context-aware intelligent systems}. In
  \bibinfo{booktitle}{\emph{Proceedings of the SIGCHI conference on human
  factors in computing systems}}. \bibinfo{pages}{2119--2128}.
\newblock


\bibitem[\protect\citeauthoryear{Linardatos, Papastefanopoulos, and
  Kotsiantis}{Linardatos et~al\mbox{.}}{2021}]%
        {survey}
\bibfield{author}{\bibinfo{person}{Pantelis Linardatos},
  \bibinfo{person}{Vasilis Papastefanopoulos}, {and}
  \bibinfo{person}{Sotiris~B. Kotsiantis}.} \bibinfo{year}{2021}\natexlab{}.
\newblock \showarticletitle{Explainable AI: A Review of Machine Learning
  Interpretability Methods}.
\newblock \bibinfo{journal}{\emph{Entropy}}  \bibinfo{volume}{23}
  (\bibinfo{year}{2021}).
\newblock


\bibitem[\protect\citeauthoryear{Lipton}{Lipton}{2018}]%
        {lipton2018mythos}
\bibfield{author}{\bibinfo{person}{Zachary~C Lipton}.}
  \bibinfo{year}{2018}\natexlab{}.
\newblock \showarticletitle{The mythos of model interpretability}.
\newblock \bibinfo{journal}{\emph{Queue}} \bibinfo{volume}{16},
  \bibinfo{number}{3} (\bibinfo{year}{2018}), \bibinfo{pages}{31--57}.
\newblock


\bibitem[\protect\citeauthoryear{Liu, Shen, Zhang, Dolan, Carin, and Chen}{Liu
  et~al\mbox{.}}{2021a}]%
        {liu2021makes}
\bibfield{author}{\bibinfo{person}{Jiachang Liu}, \bibinfo{person}{Dinghan
  Shen}, \bibinfo{person}{Yizhe Zhang}, \bibinfo{person}{Bill Dolan},
  \bibinfo{person}{Lawrence Carin}, {and} \bibinfo{person}{Weizhu Chen}.}
  \bibinfo{year}{2021}\natexlab{a}.
\newblock \showarticletitle{What Makes Good In-Context Examples for GPT-$3 $?}
\newblock \bibinfo{journal}{\emph{arXiv preprint arXiv:2101.06804}}
  (\bibinfo{year}{2021}).
\newblock


\bibitem[\protect\citeauthoryear{Liu, Yuan, Fu, Jiang, Hayashi, and Neubig}{Liu
  et~al\mbox{.}}{2021b}]%
        {liu2021pre}
\bibfield{author}{\bibinfo{person}{Pengfei Liu}, \bibinfo{person}{Weizhe Yuan},
  \bibinfo{person}{Jinlan Fu}, \bibinfo{person}{Zhengbao Jiang},
  \bibinfo{person}{Hiroaki Hayashi}, {and} \bibinfo{person}{Graham Neubig}.}
  \bibinfo{year}{2021}\natexlab{b}.
\newblock \showarticletitle{Pre-train, prompt, and predict: A systematic survey
  of prompting methods in natural language processing}.
\newblock \bibinfo{journal}{\emph{arXiv preprint arXiv:2107.13586}}
  (\bibinfo{year}{2021}).
\newblock


\bibitem[\protect\citeauthoryear{Louie, Coenen, Huang, Terry, and Cai}{Louie
  et~al\mbox{.}}{2020a}]%
        {louie2020novice}
\bibfield{author}{\bibinfo{person}{Ryan Louie}, \bibinfo{person}{Andy Coenen},
  \bibinfo{person}{Cheng~Zhi Huang}, \bibinfo{person}{Michael Terry}, {and}
  \bibinfo{person}{Carrie~J Cai}.} \bibinfo{year}{2020}\natexlab{a}.
\newblock \showarticletitle{Novice-AI music co-creation via AI-steering tools
  for deep generative models}. In \bibinfo{booktitle}{\emph{Proceedings of the
  2020 CHI Conference on Human Factors in Computing Systems}}.
  \bibinfo{pages}{1--13}.
\newblock


\bibitem[\protect\citeauthoryear{Louie, Cohen, Huang, Terry, and Cai}{Louie
  et~al\mbox{.}}{2020b}]%
        {louie2020cococo}
\bibfield{author}{\bibinfo{person}{Ryan Louie}, \bibinfo{person}{Any Cohen},
  \bibinfo{person}{Cheng-Zhi~Anna Huang}, \bibinfo{person}{Michael Terry},
  {and} \bibinfo{person}{Carrie~J Cai}.} \bibinfo{year}{2020}\natexlab{b}.
\newblock \showarticletitle{Cococo: AI-Steering Tools for Music Novices
  Co-Creating with Generative Models.}. In \bibinfo{booktitle}{\emph{HAI-GEN+
  user2agent@ IUI}}.
\newblock


\bibitem[\protect\citeauthoryear{Lu, Guo, Ren, Huang, Svyatkovskiy, Blanco,
  Clement, Drain, Jiang, Tang, Li, Zhou, Shou, Zhou, Tufano, Gong, Zhou, Duan,
  Sundaresan, Deng, Fu, and Liu}{Lu et~al\mbox{.}}{2021}]%
        {Lu2021CodeXGLUEAM}
\bibfield{author}{\bibinfo{person}{Shuai Lu}, \bibinfo{person}{Daya Guo},
  \bibinfo{person}{Shuo Ren}, \bibinfo{person}{Junjie Huang},
  \bibinfo{person}{Alexey Svyatkovskiy}, \bibinfo{person}{Ambrosio Blanco},
  \bibinfo{person}{Colin Clement}, \bibinfo{person}{Dawn Drain},
  \bibinfo{person}{Daxin Jiang}, \bibinfo{person}{Duyu Tang},
  \bibinfo{person}{Ge Li}, \bibinfo{person}{Lidong Zhou},
  \bibinfo{person}{Linjun Shou}, \bibinfo{person}{Long Zhou},
  \bibinfo{person}{Michele Tufano}, \bibinfo{person}{Ming Gong},
  \bibinfo{person}{Ming Zhou}, \bibinfo{person}{Nan Duan},
  \bibinfo{person}{Neel Sundaresan}, \bibinfo{person}{Shao~Kun Deng},
  \bibinfo{person}{Shengyu Fu}, {and} \bibinfo{person}{Shujie Liu}.}
  \bibinfo{year}{2021}\natexlab{}.
\newblock \showarticletitle{CodeXGLUE: A Machine Learning Benchmark Dataset for
  Code Understanding and Generation}.
\newblock \bibinfo{journal}{\emph{ArXiv}}  \bibinfo{volume}{abs/2102.04664}
  (\bibinfo{year}{2021}).
\newblock


\bibitem[\protect\citeauthoryear{Lundberg and Lee}{Lundberg and Lee}{2017}]%
        {lundberg2017unified}
\bibfield{author}{\bibinfo{person}{Scott~M Lundberg} {and}
  \bibinfo{person}{Su-In Lee}.} \bibinfo{year}{2017}\natexlab{}.
\newblock \showarticletitle{A unified approach to interpreting model
  predictions}. In \bibinfo{booktitle}{\emph{Proceedings of the 31st
  international conference on neural information processing systems}}.
  \bibinfo{pages}{4768--4777}.
\newblock


\bibitem[\protect\citeauthoryear{Metz}{Metz}{2021}]%
        {cade21ai}
\bibfield{author}{\bibinfo{person}{Cade Metz}.}
  \bibinfo{year}{2021}\natexlab{}.
\newblock \showarticletitle{A.I. Can Now Write Its Own Computer Code. That’s
  Good News for Humans.}
\newblock \bibinfo{journal}{\emph{The New York Times}} (\bibinfo{date}{9
  September} \bibinfo{year}{2021}).
\newblock
\urldef\tempurl%
\url{https://www.nytimes.com/2021/09/09/technology/codex-artificial-intelligence-coding.html}
\showURL{%
\tempurl}


\bibitem[\protect\citeauthoryear{Mitchell, Wu, Zaldivar, Barnes, Vasserman,
  Hutchinson, Spitzer, Raji, and Gebru}{Mitchell et~al\mbox{.}}{2019}]%
        {Mitchell2019ModelCF}
\bibfield{author}{\bibinfo{person}{Margaret Mitchell}, \bibinfo{person}{Simone
  Wu}, \bibinfo{person}{Andrew Zaldivar}, \bibinfo{person}{Parker Barnes},
  \bibinfo{person}{Lucy Vasserman}, \bibinfo{person}{Ben Hutchinson},
  \bibinfo{person}{Elena Spitzer}, \bibinfo{person}{Inioluwa~Deborah Raji},
  {and} \bibinfo{person}{Timnit Gebru}.} \bibinfo{year}{2019}\natexlab{}.
\newblock \showarticletitle{Model Cards for Model Reporting}.
\newblock \bibinfo{journal}{\emph{Proceedings of the Conference on Fairness,
  Accountability, and Transparency}} (\bibinfo{year}{2019}).
\newblock


\bibitem[\protect\citeauthoryear{Muller, Angelov, Guha, Kogan, Neff, Oliver,
  Rodriquez, and Weller}{Muller et~al\mbox{.}}{2021a}]%
        {muller2021hcai}
\bibfield{author}{\bibinfo{person}{Michael Muller}, \bibinfo{person}{Plamen
  Angelov}, \bibinfo{person}{Shion Guha}, \bibinfo{person}{Marina Kogan},
  \bibinfo{person}{Gina Neff}, \bibinfo{person}{Nuria Oliver},
  \bibinfo{person}{Manuel~Gomez Rodriquez}, {and} \bibinfo{person}{Adrian
  Weller}.} \bibinfo{year}{2021}\natexlab{a}.
\newblock \bibinfo{booktitle}{\emph{HCAI@NeurIPS2021: Human Centered AI
  workshop at NeurIPS 2021}}.
\newblock
\urldef\tempurl%
\url{https://sites.google.com/view/hcai-human-centered-ai-neurips/home}
\showURL{%
\tempurl}
\newblock
\shownote{Accessed January 17, 2022.}


\bibitem[\protect\citeauthoryear{Muller, Aragon, Guha, Kogan, Neff, Seidelin,
  Shilton, and Tanweer}{Muller et~al\mbox{.}}{2020}]%
        {muller2020interrogating}
\bibfield{author}{\bibinfo{person}{Michael Muller}, \bibinfo{person}{Cecilia
  Aragon}, \bibinfo{person}{Shion Guha}, \bibinfo{person}{Marina Kogan},
  \bibinfo{person}{Gina Neff}, \bibinfo{person}{Cathrine Seidelin},
  \bibinfo{person}{Katie Shilton}, {and} \bibinfo{person}{Anissa Tanweer}.}
  \bibinfo{year}{2020}\natexlab{}.
\newblock \showarticletitle{Interrogating Data Science}. In
  \bibinfo{booktitle}{\emph{Conference Companion Publication of the 2020 on
  Computer Supported Cooperative Work and Social Computing}}.
  \bibinfo{pages}{467--473}.
\newblock


\bibitem[\protect\citeauthoryear{Muller, Feinberg, George, Jackson, John, Kery,
  and Passi}{Muller et~al\mbox{.}}{2019}]%
        {muller2019human}
\bibfield{author}{\bibinfo{person}{Michael Muller}, \bibinfo{person}{Melanie
  Feinberg}, \bibinfo{person}{Timothy George}, \bibinfo{person}{Steven~J
  Jackson}, \bibinfo{person}{Bonnie~E John}, \bibinfo{person}{Mary~Beth Kery},
  {and} \bibinfo{person}{Samir Passi}.} \bibinfo{year}{2019}\natexlab{}.
\newblock \showarticletitle{Human-centered study of data science work
  practices}. In \bibinfo{booktitle}{\emph{Extended Abstracts of the 2019 CHI
  Conference on Human Factors in Computing Systems}}. \bibinfo{pages}{1--8}.
\newblock


\bibitem[\protect\citeauthoryear{Muller and Liao}{Muller and Liao}{[n.d.]}]%
        {muller2017exploring}
\bibfield{author}{\bibinfo{person}{Michael Muller} {and}
  \bibinfo{person}{Q~Vera Liao}.} \bibinfo{year}{[n.d.]}\natexlab{}.
\newblock \showarticletitle{Exploring AI Ethics and Values through
  Participatory Design Fictions}.
\newblock  (\bibinfo{year}{[n.\,d.]}).
\newblock


\bibitem[\protect\citeauthoryear{Muller, Wang, Ross, Weisz, Agarwal,
  Talamadupula, Houde, Martinez, Richards, Drozdal, Lui, Piorkowski, and
  Wang}{Muller et~al\mbox{.}}{2021b}]%
        {muller2021datascientists}
\bibfield{author}{\bibinfo{person}{Michael Muller}, \bibinfo{person}{April~Y.
  Wang}, \bibinfo{person}{Steven~I. Ross}, \bibinfo{person}{Justin~D. Weisz},
  \bibinfo{person}{Mayank Agarwal}, \bibinfo{person}{Kartik Talamadupula},
  \bibinfo{person}{Stephanie Houde}, \bibinfo{person}{Fernando Martinez},
  \bibinfo{person}{John Richards}, \bibinfo{person}{Jaimie Drozdal},
  \bibinfo{person}{Xie Lui}, \bibinfo{person}{David Piorkowski}, {and}
  \bibinfo{person}{Dakuo Wang}.} \bibinfo{year}{2021}\natexlab{b}.
\newblock \bibinfo{booktitle}{\emph{How data scientists improve generated code
  documentation in Jupyter notebooks}}.
\newblock
\urldef\tempurl%
\url{https://hai-gen2021.github.io/program/}
\showURL{%
Retrieved October 5, 2021 from \tempurl}


\bibitem[\protect\citeauthoryear{Muller, Wolf, Andres, Desmond, Joshi,
  Ashktorab, Sharma, Brimijoin, Pan, Duesterwald, et~al\mbox{.}}{Muller
  et~al\mbox{.}}{2021c}]%
        {muller2021designing}
\bibfield{author}{\bibinfo{person}{Michael Muller},
  \bibinfo{person}{Christine~T Wolf}, \bibinfo{person}{Josh Andres},
  \bibinfo{person}{Michael Desmond}, \bibinfo{person}{Narendra~Nath Joshi},
  \bibinfo{person}{Zahra Ashktorab}, \bibinfo{person}{Aabhas Sharma},
  \bibinfo{person}{Kristina Brimijoin}, \bibinfo{person}{Qian Pan},
  \bibinfo{person}{Evelyn Duesterwald}, {et~al\mbox{.}}}
  \bibinfo{year}{2021}\natexlab{c}.
\newblock \showarticletitle{Designing Ground Truth and the Social Life of
  Labels}. In \bibinfo{booktitle}{\emph{Proceedings of the 2021 CHI Conference
  on Human Factors in Computing Systems}}. \bibinfo{pages}{1--16}.
\newblock


\bibitem[\protect\citeauthoryear{Nguyen, Nguyen, and Nguyen}{Nguyen
  et~al\mbox{.}}{2014}]%
        {nguyen2014migrating}
\bibfield{author}{\bibinfo{person}{Anh~Tuan Nguyen},
  \bibinfo{person}{Tung~Thanh Nguyen}, {and} \bibinfo{person}{Tien~N Nguyen}.}
  \bibinfo{year}{2014}\natexlab{}.
\newblock \showarticletitle{Migrating code with statistical machine
  translation}. In \bibinfo{booktitle}{\emph{{Companion Proceedings of the 36th
  International Conference on Software Engineering}}}.
  \bibinfo{pages}{544--547}.
\newblock


\bibitem[\protect\citeauthoryear{Oda, Fudaba, Neubig, Hata, Sakti, Toda, and
  Nakamura}{Oda et~al\mbox{.}}{2015}]%
        {oda2015learning}
\bibfield{author}{\bibinfo{person}{Yusuke Oda}, \bibinfo{person}{Hiroyuki
  Fudaba}, \bibinfo{person}{Graham Neubig}, \bibinfo{person}{Hideaki Hata},
  \bibinfo{person}{Sakriani Sakti}, \bibinfo{person}{Tomoki Toda}, {and}
  \bibinfo{person}{Satoshi Nakamura}.} \bibinfo{year}{2015}\natexlab{}.
\newblock \showarticletitle{Learning to generate pseudo-code from source code
  using statistical machine translation (t)}. In
  \bibinfo{booktitle}{\emph{{2015 30th IEEE/ACM International Conference on
  Automated Software Engineering (ASE)}}}. IEEE, \bibinfo{pages}{574--584}.
\newblock


\bibitem[\protect\citeauthoryear{P{\'a}ez}{P{\'a}ez}{2019}]%
        {paez2019pragmatic}
\bibfield{author}{\bibinfo{person}{Andr{\'e}s P{\'a}ez}.}
  \bibinfo{year}{2019}\natexlab{}.
\newblock \showarticletitle{The pragmatic turn in explainable artificial
  intelligence (XAI)}.
\newblock \bibinfo{journal}{\emph{Minds and Machines}} \bibinfo{volume}{29},
  \bibinfo{number}{3} (\bibinfo{year}{2019}), \bibinfo{pages}{441--459}.
\newblock


\bibitem[\protect\citeauthoryear{Parasuraman, Sheridan, and
  Wickens}{Parasuraman et~al\mbox{.}}{2000}]%
        {parasuraman2000model}
\bibfield{author}{\bibinfo{person}{Raja Parasuraman}, \bibinfo{person}{Thomas~B
  Sheridan}, {and} \bibinfo{person}{Christopher~D Wickens}.}
  \bibinfo{year}{2000}\natexlab{}.
\newblock \showarticletitle{A model for types and levels of human interaction
  with automation}.
\newblock \bibinfo{journal}{\emph{IEEE Transactions on systems, man, and
  cybernetics-Part A: Systems and Humans}} \bibinfo{volume}{30},
  \bibinfo{number}{3} (\bibinfo{year}{2000}), \bibinfo{pages}{286--297}.
\newblock


\bibitem[\protect\citeauthoryear{Piorkowski, Gonz'alez, Richards, and
  Houde}{Piorkowski et~al\mbox{.}}{2020}]%
        {Piorkowski2020TowardsEA}
\bibfield{author}{\bibinfo{person}{David Piorkowski}, \bibinfo{person}{D.
  Gonz'alez}, \bibinfo{person}{John~T. Richards}, {and}
  \bibinfo{person}{Stephanie Houde}.} \bibinfo{year}{2020}\natexlab{}.
\newblock \showarticletitle{Towards evaluating and eliciting high-quality
  documentation for intelligent systems}.
\newblock \bibinfo{journal}{\emph{ArXiv}}  \bibinfo{volume}{abs/2011.08774}
  (\bibinfo{year}{2020}).
\newblock


\bibitem[\protect\citeauthoryear{Piorkowski, Park, Wang, Wang, Muller, and
  Portnoy}{Piorkowski et~al\mbox{.}}{2021}]%
        {Piorkowski2021HowAD}
\bibfield{author}{\bibinfo{person}{David Piorkowski}, \bibinfo{person}{Soya
  Park}, \bibinfo{person}{A. Wang}, \bibinfo{person}{Dakuo Wang},
  \bibinfo{person}{Michael~J. Muller}, {and} \bibinfo{person}{Felix Portnoy}.}
  \bibinfo{year}{2021}\natexlab{}.
\newblock \showarticletitle{How AI Developers Overcome Communication Challenges
  in a Multidisciplinary Team}.
\newblock \bibinfo{journal}{\emph{Proceedings of the ACM on Human-Computer
  Interaction}}  \bibinfo{volume}{5} (\bibinfo{year}{2021}), \bibinfo{pages}{1
  -- 25}.
\newblock


\bibitem[\protect\citeauthoryear{Puri, Kung, Janssen, Zhang, Domeniconi,
  Zolotov, Dolby, Chen, Choudhury, Decker, Thost, Buratti, Pujar, and
  Finkler}{Puri et~al\mbox{.}}{2021}]%
        {codenet}
\bibfield{author}{\bibinfo{person}{Ruchi Puri}, \bibinfo{person}{D. Kung},
  \bibinfo{person}{G. Janssen}, \bibinfo{person}{Wei Zhang},
  \bibinfo{person}{Giacomo Domeniconi}, \bibinfo{person}{Vladmir Zolotov},
  \bibinfo{person}{Julian Dolby}, \bibinfo{person}{Jie Chen},
  \bibinfo{person}{M. Choudhury}, \bibinfo{person}{Lindsey Decker},
  \bibinfo{person}{Veronika Thost}, \bibinfo{person}{Luca Buratti},
  \bibinfo{person}{Saurabh Pujar}, {and} \bibinfo{person}{Ulrich Finkler}.}
  \bibinfo{year}{2021}\natexlab{}.
\newblock \showarticletitle{Project CodeNet: A Large-Scale AI for Code Dataset
  for Learning a Diversity of Coding Tasks}.
\newblock \bibinfo{journal}{\emph{ArXiv}}  \bibinfo{volume}{abs/2105.12655}
  (\bibinfo{year}{2021}).
\newblock


\bibitem[\protect\citeauthoryear{Raji and Yang}{Raji and Yang}{2019}]%
        {raji2019ml}
\bibfield{author}{\bibinfo{person}{Inioluwa~Deborah Raji} {and}
  \bibinfo{person}{Jingying Yang}.} \bibinfo{year}{2019}\natexlab{}.
\newblock \showarticletitle{About ml: Annotation and benchmarking on
  understanding and transparency of machine learning lifecycles}.
\newblock \bibinfo{journal}{\emph{arXiv preprint arXiv:1912.06166}}
  (\bibinfo{year}{2019}).
\newblock


\bibitem[\protect\citeauthoryear{Rani}{Rani}{2017}]%
        {Rani2017ADS}
\bibfield{author}{\bibinfo{person}{Sahil Barjtya Ankur Sharma~Usha Rani}.}
  \bibinfo{year}{2017}\natexlab{}.
\newblock \showarticletitle{A detailed study of Software Development Life Cycle
  (SDLC) Models}.
\newblock \bibinfo{journal}{\emph{International Journal of Engineering and
  Computer Science}}  \bibinfo{volume}{6} (\bibinfo{year}{2017}).
\newblock


\bibitem[\protect\citeauthoryear{Ribeiro, Singh, and Guestrin}{Ribeiro
  et~al\mbox{.}}{2016}]%
        {ribeiro2016should}
\bibfield{author}{\bibinfo{person}{Marco~Tulio Ribeiro},
  \bibinfo{person}{Sameer Singh}, {and} \bibinfo{person}{Carlos Guestrin}.}
  \bibinfo{year}{2016}\natexlab{}.
\newblock \showarticletitle{Why should i trust you?: Explaining the predictions
  of any classifier}. In \bibinfo{booktitle}{\emph{Proceedings of KDD}}.
\newblock


\bibitem[\protect\citeauthoryear{Richards, Piorkowski, Hind, Houde, and
  Mojsilovi{\'c}}{Richards et~al\mbox{.}}{2020a}]%
        {richards2020methodology}
\bibfield{author}{\bibinfo{person}{John Richards}, \bibinfo{person}{David
  Piorkowski}, \bibinfo{person}{Michael Hind}, \bibinfo{person}{Stephanie
  Houde}, {and} \bibinfo{person}{Aleksandra Mojsilovi{\'c}}.}
  \bibinfo{year}{2020}\natexlab{a}.
\newblock \showarticletitle{A Methodology for Creating AI FactSheets}.
\newblock \bibinfo{journal}{\emph{arXiv preprint arXiv:2006.13796}}
  (\bibinfo{year}{2020}).
\newblock


\bibitem[\protect\citeauthoryear{Richards, Piorkowski, Hind, Houde, and
  Mojsilovi'c}{Richards et~al\mbox{.}}{2020b}]%
        {Richards2020AMF}
\bibfield{author}{\bibinfo{person}{John~T. Richards}, \bibinfo{person}{David
  Piorkowski}, \bibinfo{person}{M. Hind}, \bibinfo{person}{Stephanie Houde},
  {and} \bibinfo{person}{Aleksandra Mojsilovi'c}.}
  \bibinfo{year}{2020}\natexlab{b}.
\newblock \showarticletitle{A Methodology for Creating AI FactSheets}.
\newblock \bibinfo{journal}{\emph{ArXiv}}  \bibinfo{volume}{abs/2006.13796}
  (\bibinfo{year}{2020}).
\newblock


\bibitem[\protect\citeauthoryear{Ridgeway}{Ridgeway}{2016}]%
        {Ridgeway2016ASO}
\bibfield{author}{\bibinfo{person}{Karl Ridgeway}.}
  \bibinfo{year}{2016}\natexlab{}.
\newblock \showarticletitle{A Survey of Inductive Biases for Factorial
  Representation-Learning}.
\newblock \bibinfo{journal}{\emph{ArXiv}}  \bibinfo{volume}{abs/1612.05299}
  (\bibinfo{year}{2016}).
\newblock


\bibitem[\protect\citeauthoryear{Ridgeway and Mozer}{Ridgeway and
  Mozer}{2018}]%
        {Ridgeway2018LearningDD}
\bibfield{author}{\bibinfo{person}{Karl Ridgeway} {and}
  \bibinfo{person}{Michael~C. Mozer}.} \bibinfo{year}{2018}\natexlab{}.
\newblock \showarticletitle{Learning Deep Disentangled Embeddings with the
  F-Statistic Loss}. In \bibinfo{booktitle}{\emph{NeurIPS}}.
\newblock


\bibitem[\protect\citeauthoryear{Riedl}{Riedl}{2019}]%
        {riedl2019human}
\bibfield{author}{\bibinfo{person}{Mark~O Riedl}.}
  \bibinfo{year}{2019}\natexlab{}.
\newblock \showarticletitle{Human-centered artificial intelligence and machine
  learning}.
\newblock \bibinfo{journal}{\emph{Human Behavior and Emerging Technologies}}
  \bibinfo{volume}{1}, \bibinfo{number}{1} (\bibinfo{year}{2019}),
  \bibinfo{pages}{33--36}.
\newblock


\bibitem[\protect\citeauthoryear{Ross, Chen, Hang, Glassman, and
  Doshi-Velez}{Ross et~al\mbox{.}}{2021}]%
        {ross2021evaluating}
\bibfield{author}{\bibinfo{person}{Andrew Ross}, \bibinfo{person}{Nina Chen},
  \bibinfo{person}{Elisa~Zhao Hang}, \bibinfo{person}{Elena~L Glassman}, {and}
  \bibinfo{person}{Finale Doshi-Velez}.} \bibinfo{year}{2021}\natexlab{}.
\newblock \showarticletitle{Evaluating the Interpretability of Generative
  Models by Interactive Reconstruction}. In
  \bibinfo{booktitle}{\emph{Proceedings of the 2021 CHI Conference on Human
  Factors in Computing Systems}}. \bibinfo{pages}{1--15}.
\newblock


\bibitem[\protect\citeauthoryear{Rosson and Carroll}{Rosson and
  Carroll}{2009}]%
        {rosson2009scenario}
\bibfield{author}{\bibinfo{person}{Mary~Beth Rosson} {and}
  \bibinfo{person}{John~M Carroll}.} \bibinfo{year}{2009}\natexlab{}.
\newblock \showarticletitle{Scenario-based design}.
\newblock In \bibinfo{booktitle}{\emph{Human-computer interaction}}.
  \bibinfo{publisher}{CRC Press}, \bibinfo{pages}{161--180}.
\newblock


\bibitem[\protect\citeauthoryear{Roziere, Lachaux, Chanussot, and
  Lample}{Roziere et~al\mbox{.}}{2020}]%
        {roziere2020unsupervised}
\bibfield{author}{\bibinfo{person}{Baptiste Roziere},
  \bibinfo{person}{Marie-Anne Lachaux}, \bibinfo{person}{Lowik Chanussot},
  {and} \bibinfo{person}{Guillaume Lample}.} \bibinfo{year}{2020}\natexlab{}.
\newblock \showarticletitle{Unsupervised Translation of Programming
  Languages.}. In \bibinfo{booktitle}{\emph{NeurIPS}}.
\newblock


\bibitem[\protect\citeauthoryear{Shneiderman}{Shneiderman}{2020}]%
        {shneiderman2020bridging}
\bibfield{author}{\bibinfo{person}{Ben Shneiderman}.}
  \bibinfo{year}{2020}\natexlab{}.
\newblock \showarticletitle{Bridging the gap between ethics and practice:
  Guidelines for reliable, safe, and trustworthy Human-Centered AI systems}.
\newblock \bibinfo{journal}{\emph{ACM Transactions on Interactive Intelligent
  Systems (TiiS)}} \bibinfo{volume}{10}, \bibinfo{number}{4}
  (\bibinfo{year}{2020}), \bibinfo{pages}{1--31}.
\newblock


\bibitem[\protect\citeauthoryear{Stuart, Dabbish, Kiesler, Kinnaird, and
  Kang}{Stuart et~al\mbox{.}}{2012}]%
        {stuart2012social}
\bibfield{author}{\bibinfo{person}{H~Colleen Stuart}, \bibinfo{person}{Laura
  Dabbish}, \bibinfo{person}{Sara Kiesler}, \bibinfo{person}{Peter Kinnaird},
  {and} \bibinfo{person}{Ruogu Kang}.} \bibinfo{year}{2012}\natexlab{}.
\newblock \showarticletitle{Social transparency in networked information
  exchange: a theoretical framework}. In \bibinfo{booktitle}{\emph{Proceedings
  of the ACM 2012 conference on Computer Supported Cooperative Work}}.
  \bibinfo{pages}{451--460}.
\newblock


\bibitem[\protect\citeauthoryear{Subramonyam, Seifert, and Adar}{Subramonyam
  et~al\mbox{.}}{2021}]%
        {subramonyam2021towards}
\bibfield{author}{\bibinfo{person}{Hariharan Subramonyam},
  \bibinfo{person}{Colleen Seifert}, {and} \bibinfo{person}{Eytan Adar}.}
  \bibinfo{year}{2021}\natexlab{}.
\newblock \showarticletitle{Towards A Process Model for Co-Creating AI
  Experiences}.
\newblock \bibinfo{journal}{\emph{arXiv preprint arXiv:2104.07595}}
  (\bibinfo{year}{2021}).
\newblock


\bibitem[\protect\citeauthoryear{Talamadupula}{Talamadupula}{2021}]%
        {talamadupula2021applied}
\bibfield{author}{\bibinfo{person}{Kartik Talamadupula}.}
  \bibinfo{year}{2021}\natexlab{}.
\newblock \showarticletitle{{Applied AI Matters - AI4Code: Applying Artificial
  Intelligence to Source Code}}.
\newblock \bibinfo{journal}{\emph{{Association for Computing Machinery (ACM)
  Special Interest Group on AI (SIGAI) AI Matters}}}  \bibinfo{volume}{7}
  (\bibinfo{year}{2021}).
\newblock
Issue 1.


\bibitem[\protect\citeauthoryear{Tufano, Drain, Svyatkovskiy, Deng, and
  Sundaresan}{Tufano et~al\mbox{.}}{2020}]%
        {tufano2020unit}
\bibfield{author}{\bibinfo{person}{Michele Tufano}, \bibinfo{person}{Dawn
  Drain}, \bibinfo{person}{Alexey Svyatkovskiy}, \bibinfo{person}{Shao~Kun
  Deng}, {and} \bibinfo{person}{Neel Sundaresan}.}
  \bibinfo{year}{2020}\natexlab{}.
\newblock \showarticletitle{Unit Test Case Generation with Transformers}.
\newblock \bibinfo{journal}{\emph{arXiv preprint arXiv:2009.05617}}
  (\bibinfo{year}{2020}).
\newblock


\bibitem[\protect\citeauthoryear{Vaughan and Wallach}{Vaughan and
  Wallach}{2020}]%
        {vaughan2020human}
\bibfield{author}{\bibinfo{person}{Jennifer~Wortman Vaughan} {and}
  \bibinfo{person}{Hanna Wallach}.} \bibinfo{year}{2020}\natexlab{}.
\newblock \showarticletitle{A human-centered agenda for intelligible machine
  learning}.
\newblock \bibinfo{journal}{\emph{Machines We Trust: Getting Along with
  Artificial Intelligence}} (\bibinfo{year}{2020}).
\newblock


\bibitem[\protect\citeauthoryear{Vig}{Vig}{2019}]%
        {Vig2019AMV}
\bibfield{author}{\bibinfo{person}{Jesse Vig}.}
  \bibinfo{year}{2019}\natexlab{}.
\newblock \showarticletitle{A Multiscale Visualization of Attention in the
  Transformer Model}. In \bibinfo{booktitle}{\emph{ACL}}.
\newblock


\bibitem[\protect\citeauthoryear{Vig and Belinkov}{Vig and Belinkov}{2019}]%
        {Vig2019AnalyzingTS}
\bibfield{author}{\bibinfo{person}{Jesse Vig} {and} \bibinfo{person}{Yonatan
  Belinkov}.} \bibinfo{year}{2019}\natexlab{}.
\newblock \showarticletitle{Analyzing the Structure of Attention in a
  Transformer Language Model}. In \bibinfo{booktitle}{\emph{BlackboxNLP@ACL}}.
\newblock


\bibitem[\protect\citeauthoryear{Vinodkumar Prabhakaran~Jr}{Vinodkumar
  Prabhakaran~Jr}{2020}]%
        {vinodkumar2020participatory}
\bibfield{author}{\bibinfo{person}{Donald~Martin Vinodkumar Prabhakaran~Jr}.}
  \bibinfo{year}{2020}\natexlab{}.
\newblock \showarticletitle{Participatory Machine Learning Using
  Community-Based System Dynamics}.
\newblock \bibinfo{journal}{\emph{Health and Human Rights}}
  \bibinfo{volume}{22}, \bibinfo{number}{2} (\bibinfo{year}{2020}),
  \bibinfo{pages}{71}.
\newblock


\bibitem[\protect\citeauthoryear{Wadhwani and Jain}{Wadhwani and Jain}{2020}]%
        {wadhwani2020machine}
\bibfield{author}{\bibinfo{person}{Abhishek Wadhwani} {and}
  \bibinfo{person}{Priyank Jain}.} \bibinfo{year}{2020}\natexlab{}.
\newblock \showarticletitle{Machine Learning Model Cards Transparency Review:
  Using model card toolkit}. In \bibinfo{booktitle}{\emph{2020 IEEE Pune
  Section International Conference (PuneCon)}}. IEEE,
  \bibinfo{pages}{133--137}.
\newblock


\bibitem[\protect\citeauthoryear{Wang, Wang, Joty, and Hoi}{Wang
  et~al\mbox{.}}{2021}]%
        {Wang2021CodeT5IU}
\bibfield{author}{\bibinfo{person}{Yue Wang}, \bibinfo{person}{Weishi Wang},
  \bibinfo{person}{Shafiq~R. Joty}, {and} \bibinfo{person}{S. Hoi}.}
  \bibinfo{year}{2021}\natexlab{}.
\newblock \showarticletitle{CodeT5: Identifier-aware Unified Pre-trained
  Encoder-Decoder Models for Code Understanding and Generation}.
\newblock


\bibitem[\protect\citeauthoryear{Weisz, Muller, Houde, Richards, Ross,
  Martinez, Agarwal, and Talamadupula}{Weisz et~al\mbox{.}}{2021}]%
        {weisz2021perfection}
\bibfield{author}{\bibinfo{person}{Justin~D Weisz}, \bibinfo{person}{Michael
  Muller}, \bibinfo{person}{Stephanie Houde}, \bibinfo{person}{John Richards},
  \bibinfo{person}{Steven~I Ross}, \bibinfo{person}{Fernando Martinez},
  \bibinfo{person}{Mayank Agarwal}, {and} \bibinfo{person}{Kartik
  Talamadupula}.} \bibinfo{year}{2021}\natexlab{}.
\newblock \showarticletitle{Perfection Not Required? Human-AI Partnerships in
  Code Translation}. In \bibinfo{booktitle}{\emph{26th International Conference
  on Intelligent User Interfaces}}. \bibinfo{pages}{402--412}.
\newblock


\bibitem[\protect\citeauthoryear{Wiegreffe and Pinter}{Wiegreffe and
  Pinter}{2019}]%
        {wiegreffe2019attention}
\bibfield{author}{\bibinfo{person}{Sarah Wiegreffe} {and}
  \bibinfo{person}{Yuval Pinter}.} \bibinfo{year}{2019}\natexlab{}.
\newblock \showarticletitle{Attention is not not explanation}.
\newblock \bibinfo{journal}{\emph{arXiv preprint arXiv:1908.04626}}
  (\bibinfo{year}{2019}).
\newblock


\bibitem[\protect\citeauthoryear{Wolf}{Wolf}{2019}]%
        {wolf2019explainability}
\bibfield{author}{\bibinfo{person}{Christine~T Wolf}.}
  \bibinfo{year}{2019}\natexlab{}.
\newblock \showarticletitle{Explainability scenarios: towards scenario-based
  XAI design}. In \bibinfo{booktitle}{\emph{Proceedings of the 24th
  International Conference on Intelligent User Interfaces}}.
  \bibinfo{pages}{252--257}.
\newblock


\bibitem[\protect\citeauthoryear{Zhang and Banovic}{Zhang and Banovic}{2021}]%
        {zhang2021method}
\bibfield{author}{\bibinfo{person}{Enhao Zhang} {and} \bibinfo{person}{Nikola
  Banovic}.} \bibinfo{year}{2021}\natexlab{}.
\newblock \showarticletitle{Method for Exploring Generative Adversarial
  Networks (GANs) via Automatically Generated Image Galleries}. In
  \bibinfo{booktitle}{\emph{Proceedings of the 2021 CHI Conference on Human
  Factors in Computing Systems}}. \bibinfo{pages}{1--15}.
\newblock


\bibitem[\protect\citeauthoryear{Zhou, Khosla, Lapedriza, Oliva, and
  Torralba}{Zhou et~al\mbox{.}}{2016}]%
        {Zhou2016LearningDF}
\bibfield{author}{\bibinfo{person}{Bolei Zhou}, \bibinfo{person}{Aditya
  Khosla}, \bibinfo{person}{{\`A}gata Lapedriza}, \bibinfo{person}{Aude Oliva},
  {and} \bibinfo{person}{Antonio Torralba}.} \bibinfo{year}{2016}\natexlab{}.
\newblock \showarticletitle{Learning Deep Features for Discriminative
  Localization}.
\newblock \bibinfo{journal}{\emph{2016 IEEE Conference on Computer Vision and
  Pattern Recognition (CVPR)}} (\bibinfo{year}{2016}),
  \bibinfo{pages}{2921--2929}.
\newblock


\bibitem[\protect\citeauthoryear{Zhu, Yu, Halfaker, and Terveen}{Zhu
  et~al\mbox{.}}{2018}]%
        {zhu2018value}
\bibfield{author}{\bibinfo{person}{Haiyi Zhu}, \bibinfo{person}{Bowen Yu},
  \bibinfo{person}{Aaron Halfaker}, {and} \bibinfo{person}{Loren Terveen}.}
  \bibinfo{year}{2018}\natexlab{}.
\newblock \showarticletitle{Value-sensitive algorithm design: Method, case
  study, and lessons}.
\newblock \bibinfo{journal}{\emph{Proceedings of the ACM on Human-Computer
  Interaction}} \bibinfo{volume}{2}, \bibinfo{number}{CSCW}
  (\bibinfo{year}{2018}), \bibinfo{pages}{1--23}.
\newblock


\end{thebibliography}

\newpage
\appendix

\section{The other two use cases}
\label{app:other-two}
\begin{figure}[tp]
    \centering
    \includegraphics[width=\linewidth]{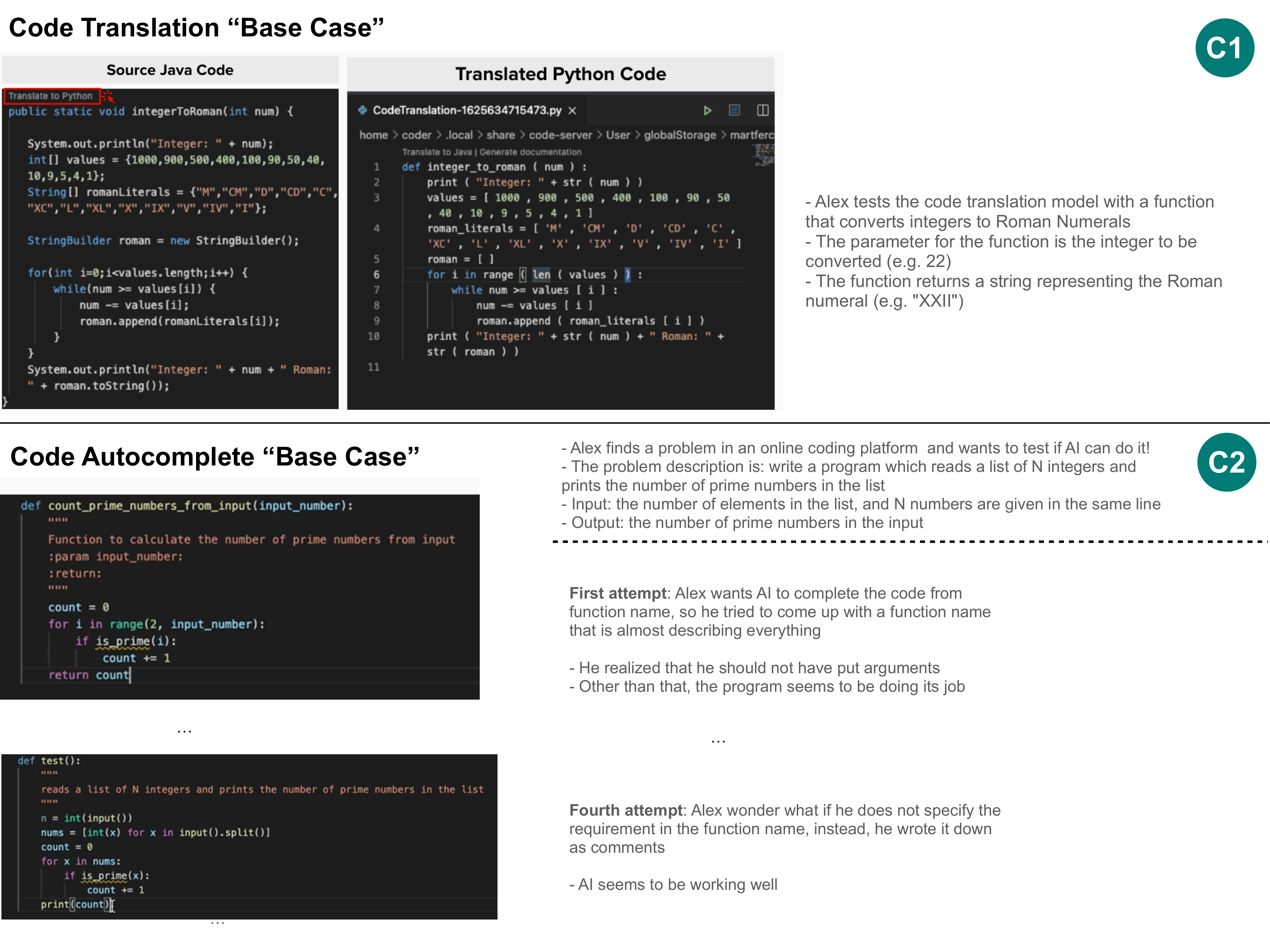}
    \caption{The two other code "base cases" we used in the workshop on Mural. C1 is the example we used as the base case for the workshops about code translation (W1-CT, W2-CT and W5-CT), and C2 is for the code autocompletion use case (W3-CA, W4-CA and W9-CA). The C in Figure~\ref{fig:workshop} is the base code base for the natural language to code use case (W6-NL2Code, W7-NL2Code, W8-NL2Code). }
    \label{fig:other-two-use-cases}
\end{figure}

We show the example code as the base case for workshops about natural language to code (W6-NL2Code, W7-NL2Code, W8-NL2Code) in Figure~\ref{fig:workshop}. As a supplement, we show how we replace Figure~\ref{fig:workshop} (c) with other code examples in Figure~\ref{fig:other-two-use-cases} for the code translation and code autocompletion use cases. The other sub-figures in Figure~\ref{fig:workshop} stay almost the same for all workshops across all use cases with minimum edits. 








\end{document}